\documentclass
[prb,aps,twocolumn,superscriptaddress,showpacs,amsmath,amssymb]{revtex4}
\usepackage{exscale,epsfig,psfrag,float}
\usepackage{color}
\usepackage{dcolumn}
\usepackage{amsmath}
\usepackage{graphicx}
\usepackage{rotating}
\usepackage[colorlinks=true,linkcolor=blue, citecolor=blue]{hyperref} 

\begin{document}

\title{Formation of magnetic moments and resistance upturn\\
        at grain boundaries of two-dimensional electron systems}

\author{Iris~Xhango}
\altaffiliation{Iris.Xhango@physik.uni-augsburg.de} 
\affiliation{Center for Electronic Correlations and Magnetism, EP VI, Institute of Physics, University of Augsburg, 86135 Augsburg, Germany}
\affiliation{Max Planck Institute for Solid State Research, Heisenbergstra{\ss}e 1, 70569 Stuttgart, Germany}
\author{Thilo Kopp}
\affiliation{Center for Electronic Correlations and Magnetism, EP VI, Institute of Physics, University of Augsburg, 86135 Augsburg, Germany}

\begin{abstract}
Electronic correlations control the normal state of bulk high-$T_c$ cuprates. Strong correlations also suppress the charge transport through cuprate grain boundaries (GBs). The question then arises if these correlations can produce magnetic states at cuprate GBs. We analyze the formation of local magnetic moments at the GB of a correlated two-dimensional electron systems which is represented by an inhomogeneous Hubbard model. The model Hamiltonian is diagonalized after the implementation of a mean-field decoupling.  The formation of local magnetic moments is supported by a sufficiently strong variance in the bond kinetic energies at the GB. Local scattering potentials can assist or suppress the formation of a magnetic GB state, depending on the details of their spacial distribution. Grain boundary induced stripes are formed in the vicinity the GB and decay into the bulk. Moreover, we observe the build-up of conducting channels which are confined by magnetic clusters. The grain boundary resistance increases at decreasing temperatures. This low-temperature behaviour is caused by the suppression of current correlations in the state with local magnetic GB moments. The resistance upturn at low temperatures is in qualitative agreement with experiments. 
\end{abstract}


\date{\today}

\maketitle

\section{Introduction}

Interfaces of high-temperature copper oxide superconductors have been in the focus of intensive experimental and theoretical research for more than 25~years (see the extended reviews, for example Refs.~\onlinecite{Hilgenkamp,Bozovic}, and references therein). In this area of research application-oriented aspects as well as fundamental theoretical issues are concerned and related in an intriguing way. For instance, cuprate grain boundaries display an exponential suppression of the critical current with increasing misalignment angle between the grains~\cite{Dimos,Hilgenkamp}. This behavior is of considerable importance for the determination of the supercurrent through Josephson junctions and the design of superconducting cables. Moreover, a detailed theoretical understanding allows to identify the nature of the reconstructed electronic states at these grain boundary interfaces and to make reliable predictions on the charge accumulated at the interface, on the formation of magnetic moments, and on the distribution of current densities through a grain boundary. 

The exponential suppression of the supercurrent~\cite{Dimos,Hilgenkamp} is related to static charge fluctuations along the grain boundary,~\cite{Graser} and the magnitude of the suppression is controlled by electronic correlations.~\cite{Wolf} The charge fluctuations originate from potential fluctuations and a distribution of bond kinetic energies, both of them produced by dislocation cores and a non-stoichiometric composition of the grain boundary. The charge profile across the grain boundary is dependent on the misalignment angle.~\cite{Graser} Large angle grain boundaries always allow for narrow streaks in the charge profile where filling is close to one hole per copper site. There, electronic correlations are most effective and suppress transport through the grain boundary which explains the observed order of magnitude of the exponential suppression.~\cite{Wolf}

Strong electronic correlations in the bulk cuprates are responsible for antiferromagnetism at and close to half-filling. Consequently, it suggests itself that cuprate interfaces and grain boundaries are also affected by strong electronic correlations~\cite{Freericks,FreericksB} and display magnetism~\cite{Andersen}, or are related to nanoscopic phase separation~\cite{Dagotto}. It is difficult to verify grain boundary magnetism directly. However, the observed linear increase of the grain boundary resistance with decreasing temperature~\cite{Schneider,Blamire} has been tentatively related to the formation of local moments~\cite{Schneider}. On the other hand, non-magnetic impurities in bulk cuprates are well known to generate magnetic moments (see Refs.~\onlinecite{Alloul,Christensen,Harter,Andersen1}, and references therein). Yet grain boundaries are extended inhomogeneities, and the electronic interface state may depend on the respective properties of the rather one-dimensional grain boundary structure.

In this work we investigate conditions on the microscopic grain boundary (GB) set-up that are favorable for the formation of magnetic moments along the GB, then present the pattern of charge currents through the GB, and analyze the temperature dependence of the GB resistance.
Actually, the properties of cuprate GBs~\cite{Schneider,Blamire,Ransley1,Ransley2,Browning,Pennycook} at elevated temperatures above $T_c$ have not been investigated so intensively but it is in this regime that magnetic moments possibly form.

Cuprate grain boundaries are characterized either as small or large angle GBs: Small angle GBs with misorientation angles up to 10$^{\circ}$ display a periodic series of dislocations to match the two lattices which are joined at the GB. In the framework of continuous elasticity theory, Gurevich and Pashitskii~\cite{Gurevich} modeled the dislocation cores as insulating, antiferromagnetic regions and explain the suppression of the critical current with increasing angle. The insulating core regions naturally provide a strong barrier for current flow and thereby produce current channels between the cores. However, the cores start to overlap beyond approximately 10$^{\circ}$, and the model does not apply in the large angle regime. There, a notable atomic scale reconstruction takes place to release strain, and a microscopic description is necessary. A molecular dynamics approach\cite{Graser}  identifies well the structural units of atomic configurations along the GB and allows to set up a microscopic modelling of the electronic phase in the presence of a GB. The exponential dependence of the critical current on the GB angle for large angle GBs has been determined within such a microscopic approach. The possible formation of GB magnetic moments has, to our knowledge, not been investigated microscopically.

To induce magnetism on the GB, we refer to a scenario where the local kinetic energies (hopping matrix elements $t$ in the bulk) are homogeneously reduced along the GB with respect to their bulk values. Assuming that the local Coulomb interaction $U$ between charge carriers is the same at the GB and in the bulk, then a reduced value of the ratio of $t/U$ at the GB may well control interfacial magnetism sufficiently close to half-filling. However, neither the hopping matrix elements nor the local potential scatterers are homogeneously distributed along the GB, and the formation of an inhomogeneous state needs a more thorough analysis.

In the first part of this article (Secs.~\ref{sec:model} and \ref{sec:magnetism}), we investigate the formation of magnetic moments at the GB when varying the bond kinetic energies and potential scattering amplitudes in the structural units that are present in a large angle GB. In the second part (Sec.~\ref{sec:transport}) we discuss the build-up of conducting channels through the GB in the presence of a magnetic interface state and the implications for the temperature dependence of the GB resistance (several details are investigated in Ref.~\onlinecite{Xhango}).

\section{Grain boundary model}\label{sec:model}

To assess the importance of electronic correlations on the GB state we model the GB with an inhomogeneous one-band Hubbard model with distinct hopping matrix elements at each bond and local potential scatterers, which parameterize the charge variations in distorted CuO$_2$ plaquettes. The on-site Coulomb interaction $U$ is approximately independent of the site although unequal screening through the neighboring O-sites may in principle modify $U$ inhomogeneously. This latter effect is neglected in our set-up. The projection onto a one-band model is a simplification which is valid if the energy scales for interband transitions are large with respect to the excitation energies at the GB. For strong potential scatterers this is  not necessarily the case. However, we emphasize that in  previous work the projection onto the one-band model produced excellent results for the dependence of the critical current on the misalignment angle.~\cite{Graser,Wolf} These results suggest that interband excitations still have sufficiently low weight to contribute significantly. In this paper we do not elaborate further on the corrections from multiband behavior but still consider it a valid concern to be investigated in the future.

The inhomogeneous one-band Hubbard model with potential scatterers parameterized by $V_i$ is:
\begin{equation} 
\label{eq:hamiltonian}
H= -\!\!\!\sum_{\langle i,j\rangle, \,\sigma} t_{ij}\, c^\dagger_{i\sigma}c^{\phantom{\dagger}}_{j\sigma} + U\,\sum_i {\hat n}_{i\uparrow}{\hat n}_{i\downarrow} +\sum_i (V_i -\mu)\,{\hat n}_i
\end{equation}
where ${\hat n}_i = \sum_{\sigma} {\hat n}_{i,\sigma}=\sum_{\sigma}c^\dagger_{i\sigma}c^{\phantom{\dagger}}_{i\sigma}$. 

\begin{figure}[tbp]
\centering
\includegraphics*[width=8.7cm]{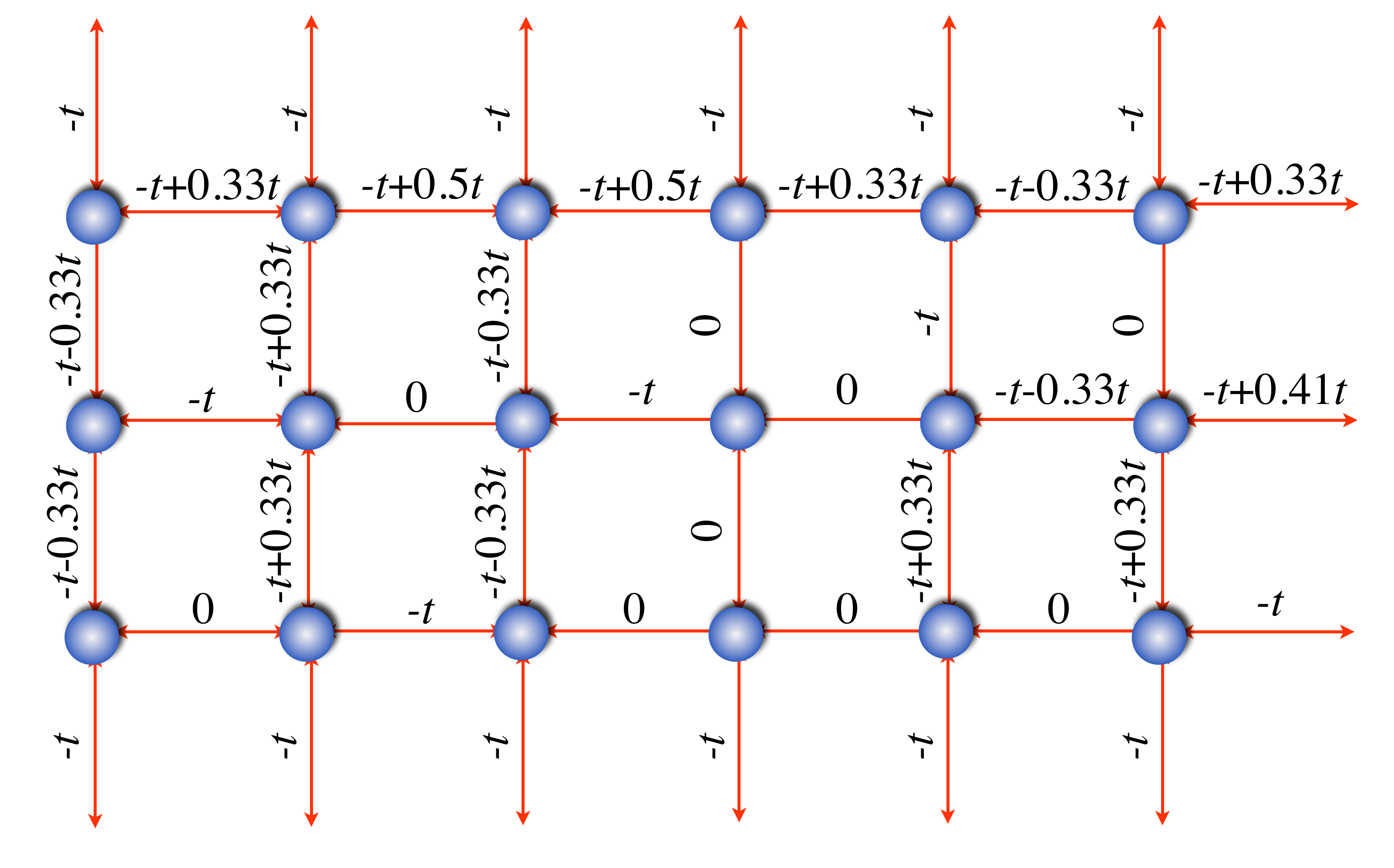}
\caption[]{(Color online) Distribution of hopping matrix elements along the GB. The GB bonds are given by the two inner vertical bonds and the horizontal bonds within the three lines of GB sites.\label{fig:GBhopping}}
\end{figure}

In this work we will focus exclusively on large-angle tilt GBs where a sequence of structural units constitutes the GB:~\cite{Browning,Pennycook} atomic patterns are repeated quasi periodically along the GB. We define the GB through the hopping matrix elements $t_{ij}$ for the bonds (see Fig.~\ref{fig:GBhopping}) and local potentials $V_{i}$ 
(see Figs.~\ref{fig:V1V2} and \ref{fig:V1V2V3}) for periodically repeated structural units of three sites width (perpendicular to the GB) and a length of six sites (along the GB). The distribution of hopping matrix elements $t_{ij}$ for the bonds within the GB are approximately adjusted to tilt GBs with misalignment angles of 30$^{\circ}$ (cf.~Ref.~\onlinecite{Graser}). In our model set-up the coordination number for sites within the GB is always four. This assumption simplifies the evaluation but does not modify our results on GB magnetism that we want to discuss rather qualitatively.

The electron-electron interaction is taken into account only through the on-site Coulomb interaction $U$ which allows to discuss the emergence of GB magnetism in the mean-field evaluation of the model Hamiltonian (\ref{eq:hamiltonian}). If not otherwise specified, we take $U=2t$, with $t$ the bulk hopping value, and adjust the chemical potential $\mu$ so that the average value of the electronic density is fixed to $n= 0.86$. The value of $U$ is chosen rather moderate in order to keep the (inhomogeneous) mean-field evaluation controlled and to prevent the system to be overly biased towards a magnetic state. The mean-field Hamiltonian for the GB model is
\begin{align} 
\label{eq:MFhamiltonian}
H_{\rm MF}\!=\! -\!\!\!\sum_{\langle i,j\rangle, \,\sigma}\!\! t_{ij}\, c^\dagger_{i\sigma}c^{\phantom{\dagger}}_{j\sigma} \!+ \!
\sum_{i, \,\sigma} \biggl[&\frac{U}{2} (n_i \!- \!\sigma m_i)+V_i \!-\!\mu)
\,c^\dagger_{i\sigma}c^{\phantom{\dagger}}_{i\sigma}\biggr]\nonumber\\
&- \,\sum_i \frac{U}{4} (n_i^2- m_i^2)
\end{align}
where $n_i= \sum_{\sigma} \langle {\hat n}_{i,\sigma}\rangle$ and $m_i= \sum_{\sigma} \sigma  \langle{\hat n}_{i,\sigma} \rangle$ are the local expectation values of electron density and magnetic moment, respectively. Both depend on the temperature $T$; we set $k_B =1$ in this work.
In the following section, we present the results for the diagonalization of this mean-field Hamiltonian on a $42 \times 20$ site lattice, with 20 sites and open boundary conditions in the direction perpendicular to the GB and 42 sites and periodic boundary conditions in the direction parallel to the GB. Larger systems (such as $60 \times 40$ sites) have been tested to confirm the convergence of the presented results.

\section{Magnetic states at the GB}\label{sec:magnetism}

\begin{figure}[H]
\begin{center}
\includegraphics*[width=8.5cm]{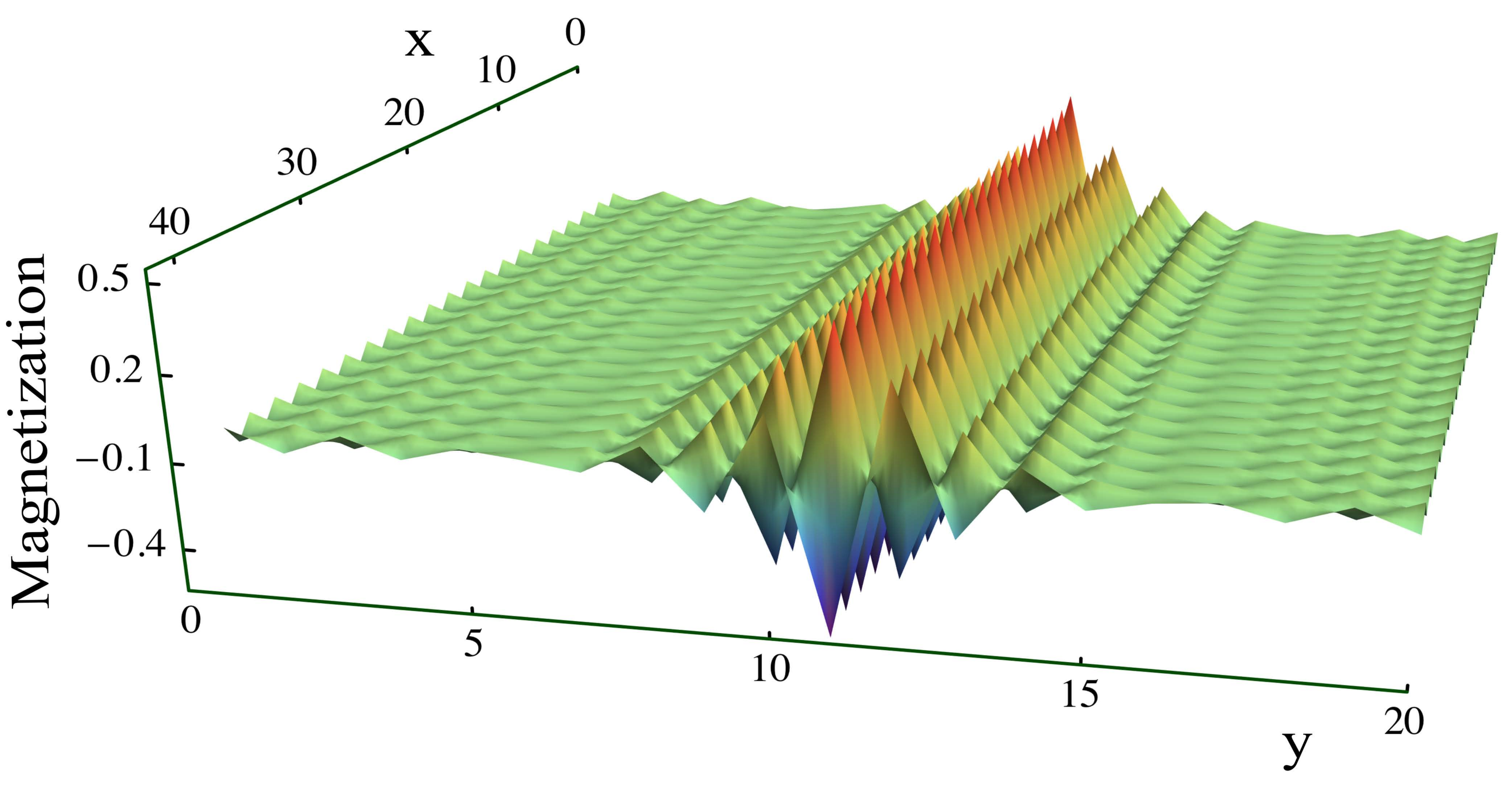}
\caption{(Color online) Site-dependent magnetization $m_i$ for a homogeneous GB: $t_{ij}=0.5\,t$ for two bonds across the boundary. The on-site Coulomb interaction is globally $U=2t$, the temperature is set to $T=0.11\,t$, and $\mu=-0.6\,t$ fixes a filling of $n=0.86$. The GB is in lines 9--11.}
\label{fig:GBhomogen}
\end{center}
\end{figure}

As preliminary test we perform a diagonalization of $H_{\rm MF}$ with $V_i =0$ and $t_{ij} = 0.5 t$ at all GB bonds of Fig.~\ref{fig:GBhopping} and determine the self-consistent solution. The values of $U=2t$ and $n=0.86$ are sufficiently high and close to half-filling, respectively, that an antiferromagnetic magnetization pattern is generated at the GB (see Fig.~\ref{fig:GBhomogen}). In fact, the magnetic state extends laterally into the bulk on further three atomic sites off the GB which indicates the non-local character of the magnetic correlations. This observation will be readdressed below.

\begin{figure}[t]
\begin{center}
\includegraphics*[width=8.5cm]{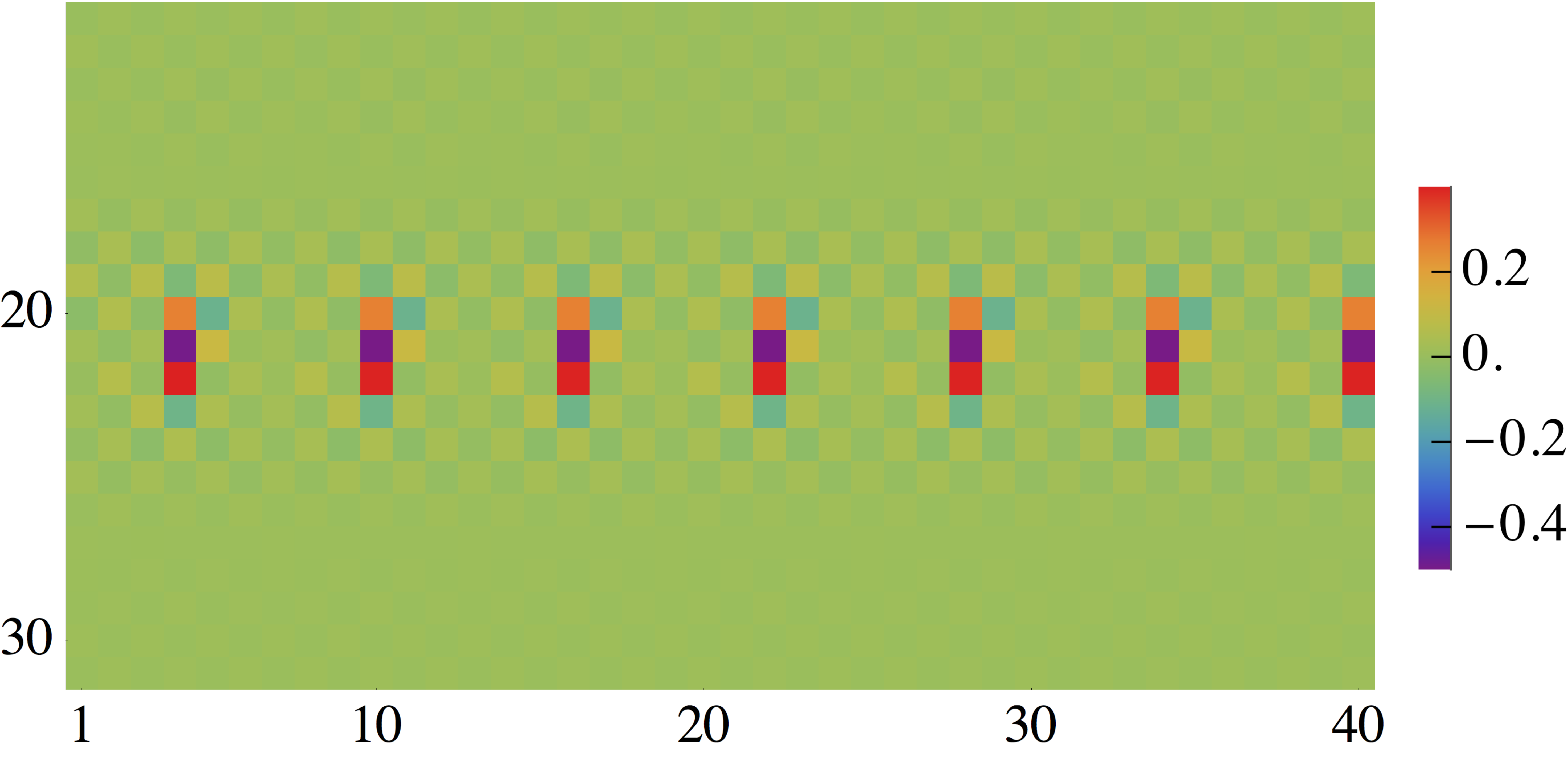}
\bigskip

\includegraphics*[width=8.5cm]{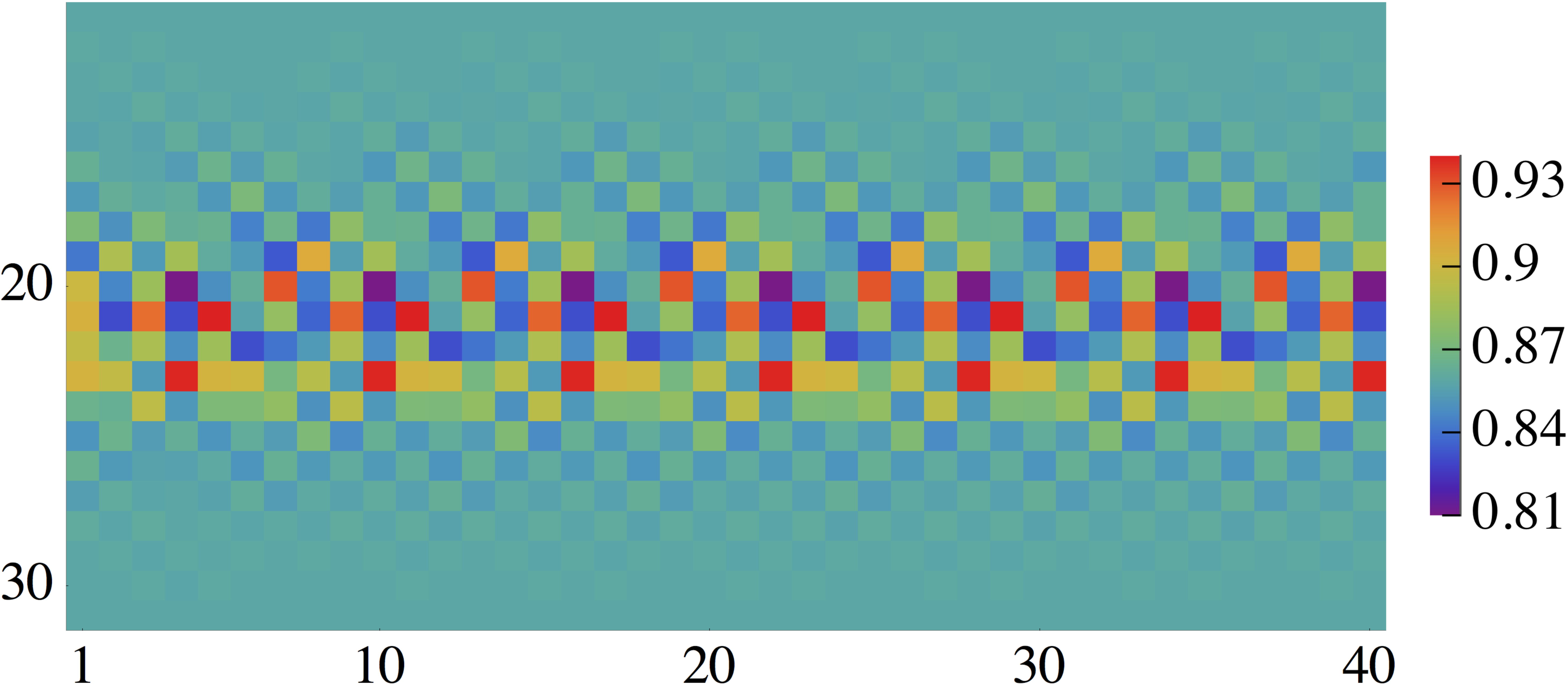}
\end{center}
\caption{(Color online) Local magnetization $m_i$ (upper panel) and electron occupation number $n_i$ (lower panel) at the GB defined through Fig.~\ref{fig:GBhopping} for a (60$\times$40) site system with periodic and open boundary conditions in the horizontal and vertical directions, respectively. The control parameters are $U=2t$, $T=0.11\,t$, and the filling is $n=0.86$. The GB is confined to lines 20--22. The figure does not display the entire  (60$\times$40) system.}
\label{fig:GBstandard}
\end{figure}

Next we investigate the relevance of a distribution of bond kinetic energies on the formation of the magnetic state. Exemplarily, we take the distribution of hopping matrix elements, which is depicted in Fig.~\ref{fig:GBhopping}.ÊAgain sizable magnetic moments are formed at the GB and decaying magnetic oscillations are seen in the nearby bulk regions (Fig.~\ref{fig:GBstandard}). This build-up of magnetic moments at the GB is expected in view of the previous result. The moments are strongest where the bonds to neighboring sites (parameterized by $t_{ij}$) are weakest. The rather antiferromagnetic character of the moment alignment in each structural unit reflects the gain in kinetic energy of antiparallel with respect to parallel alignment. The electronic density, that is, the expectation value of the site occupation also varies in the GB region: it is largest in the middle of the GB where the magnetization is highest (see Fig.~\ref{fig:GBstandard}b). Such an association is anticipated from the homogeneous mean-field solution. However, the variations are limited to a range of 0.81 to 0.93, and the highest occupation is not necessarily on the sites with largest magnetic moment.

More compelling is the question if magnetic moments can be formed through the variation of bond kinetic energies within the GB. The essential issue is if disorder within the interface can generate magnetism for given Coulomb interaction strength. In fact, from Table~\ref{Tab1} one learns that increasing the variance  $\langle\Delta t\rangle$ of the bond kinetic energies induces a transition from a non-magnetic state at finite temperature $T=0.11\,t$ to a state with robust magnetic moments for $\langle\Delta t\rangle_c \gtrsim 0.5\,t$ at constant temperature and approximately the same average bond energy $\langle t\rangle = 0.65\,t$. Here, the maximal magnetic moment in a structural GB unit is  identified from $m= \max_i |m_i|$ where $i$ is a site in the periodically repeated structural unit along the GB. Average and variance are taken from sums over the bonds of a structural unit:
$\langle t\rangle = 1/N_b\, \sum_{\langle i, j \rangle} t_{ij}$ and  
$\langle \Delta t\rangle = 1/N_b\, [\sum_{\langle i, j \rangle} t_{ij}^2]^{\frac{1}{2}}$, where the sum runs over the $N_b$ bonds of the structural unit at the GB. The average magnetic moments in the systems with a variance $\langle\Delta t\rangle$ larger than the critical variance $\langle\Delta t\rangle_c$ are approximately independent of $\langle\Delta t\rangle$. However, they increase with increasing $U$.

\begin{table}[t]
\begin{tabular}[c]{|c|c|c|c|c|c|}\hline
$<t>$ & -0.64& -0.64&-0.69 &-0.66&  -0.67  \\ \hline
$\Delta t$&\phantom{-}0.36 & \phantom{-}0.40&\phantom{-}0.50 & \phantom{-}0.57 & \phantom{-}0.59   \\ \hline\hline
$m$ & 0 & 0&\phantom{-}0.5 & \phantom{-}0.4 & \phantom{-}0.4   \\ \hline
\end{tabular}
\caption{GB magnetization $m= \max_i |m_i|$ in dependence on the variance $\langle\Delta t\rangle$ of the hopping amplitudes $t_{ij}$ within a GB structural unit. Different configurations with approximately equal $\langle t\rangle$ have been evaluated for $T=0.11\,t$ and $U=2t$.  The GB quantities $\langle\Delta t\rangle$ and $\langle t\rangle$ are in units of the bulk $t$. }
\label{Tab1}
\end{table}

Eventually, we introduce on-site scattering potentials $V_i$ to examine their impact on the magnetic state. They arise from a non-stoichiometric composition of the structural units at the GB and may act repulsively (positive potential) or attractively (negative) for electronic GB states; the latter is, for example, the case for missing oxygen ions (vacancies). It is straightforward to include these local potential scatterers in the diagonalization. Weak scattering with $|V_i| < U$ does not modify the magnetic state significantly. Here, we consider rather strong scatteres with $|V_i| >U$, viz. $|V_i|= 10t$ and $20t$ which is in the same range as the scattering potentials identified in Ref.~\onlinecite{Graser} for cuprate large-angle GBs.  For the analysis of the magnetic state, the sign is not relevant in the case of a strong local potential: a positive potential produces a nearly empty site whereas a negative potential attracts two electrons and generates a doubly occupied site. In both cases the site is nonmagnetic.

\begin{figure}[t]
(a) 
 
 \includegraphics[width=4.5cm]{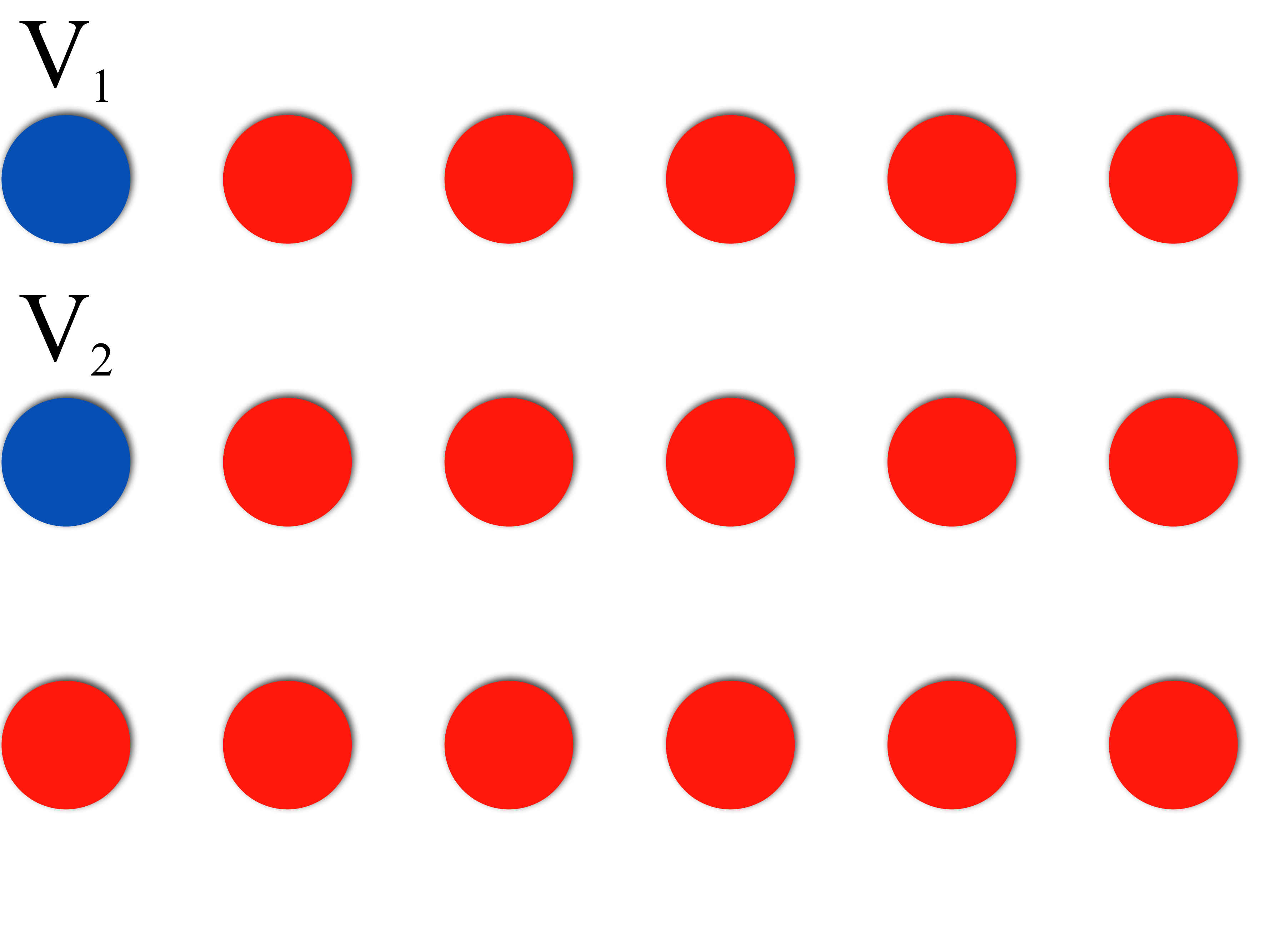}
\smallskip

(b)
\smallskip

\includegraphics[width=8.8cm]{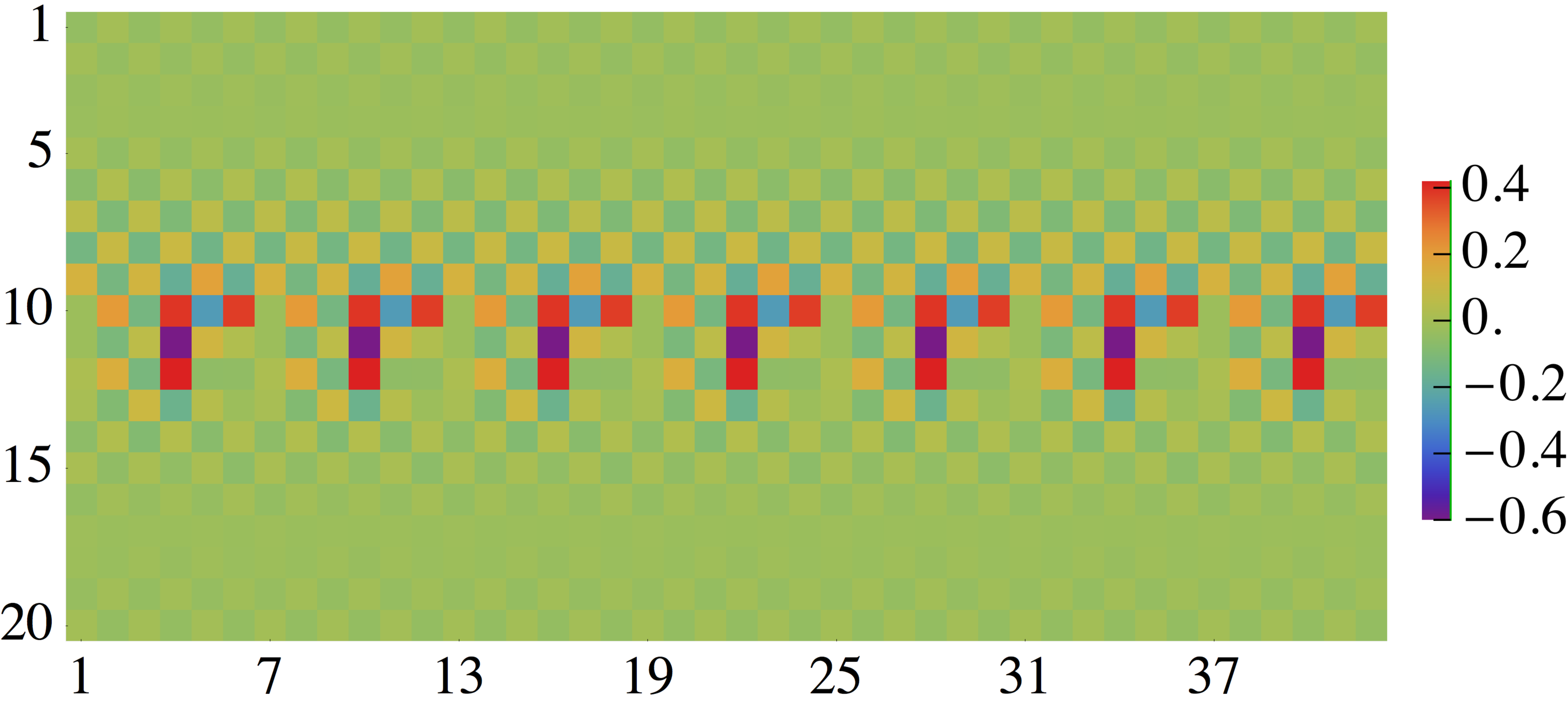}  
\smallskip

(c)
\smallskip

\includegraphics[width=8.6cm]{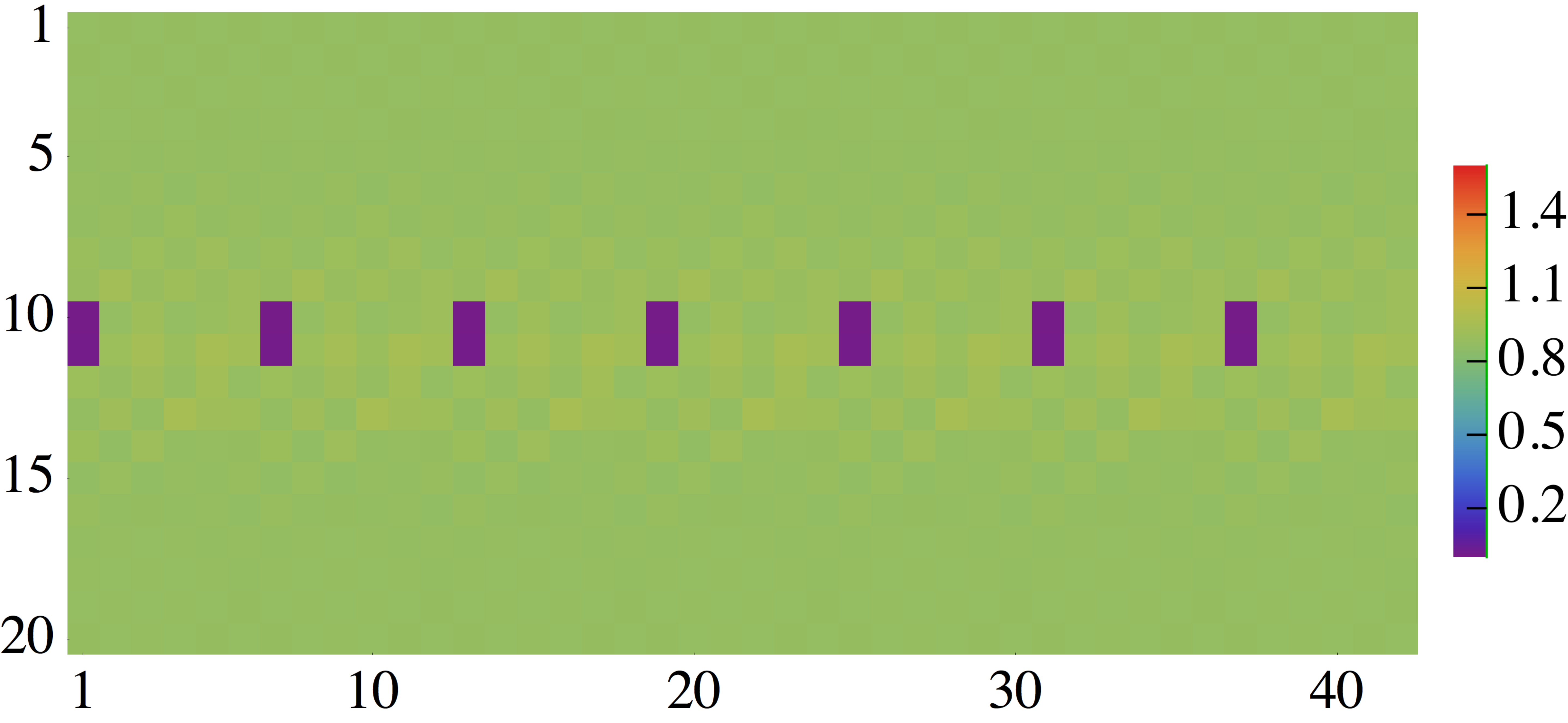} \hfill 
\caption{(Color online) GB with hopping amplitudes from Fig.~\ref{fig:GBhopping} and potential scatterers $V_{1,2}$.  (a) Scheme for the potential scatterers in a GB structural unit. (b) Local magnetization $m_i$ and (c) electron occupation number $n_i$ for a (42$\times$20) site system. The scheme from (a) translates into a potential $V_1$ at each first site (modulo 6) of line ten and $V_2$ correspondingly in line 11. Here $V_{1,2}=20\,t$, $U=2t$, $T=0.11\,t$, and the filling has been fixed to $n=0.86$.}
\label{fig:V1V2}
\end{figure}

We distinguish two scenarios: assisted magnetization and suppression of local moments. Obviously, the sites with strong potentials do not allow for the formation of local magnetic moments. This is confirmed by the magnetization patterns in Fig.~\ref{fig:V1V2}b. For positive local potentials $V_{1,2}$ (see Fig.~\ref{fig:V1V2}a for the assignment of the sites), we identify them as empty sites (Fig.~\ref{fig:V1V2}c) 
where the magnetization is zero (Fig.~\ref{fig:V1V2}b). The three sites across the GB that carry the strongest magnetization display nearly unaltered magnetic moments when the potentials are set (cf.\ Fig.~\ref{fig:GBstandard}a for $V_i=0$ and Fig.~\ref{fig:V1V2}b for $V_{1,2}=20\,t$). Surprisingly though, other sites carry a stronger magnetic moment for finite potential---most pronounced is the increase of magnetic moment on the site which is on the left to the scatterer in line~10. The reason for this increase is the additional inhomogeneity which is introduced through the potential scatterers. In particular, the addressed site (line 10, row 4 modulo 6) suffers a decrease of bond kinetic energy as the nearest-neighbor site to the right is constrained to stay empty on account of the strong positive potential. In this scenario, the potential scatterers of either sign assist the build-up of magnetism.
We note that placing a strong potential scatterer on a site that had a large magnetic moment in the absence of the scatterer---for example, on the site in line~11 and row~4 which carries the strongest moment in Fig.~\ref{fig:V1V2}b---quenches the moment on this site but does not suppress magnetism substantially at other sites. In this respect, GB magnetism appears to be robust.

\begin{figure}[t]
\begin{center}
(a) 
 
\includegraphics[width=4.5cm]{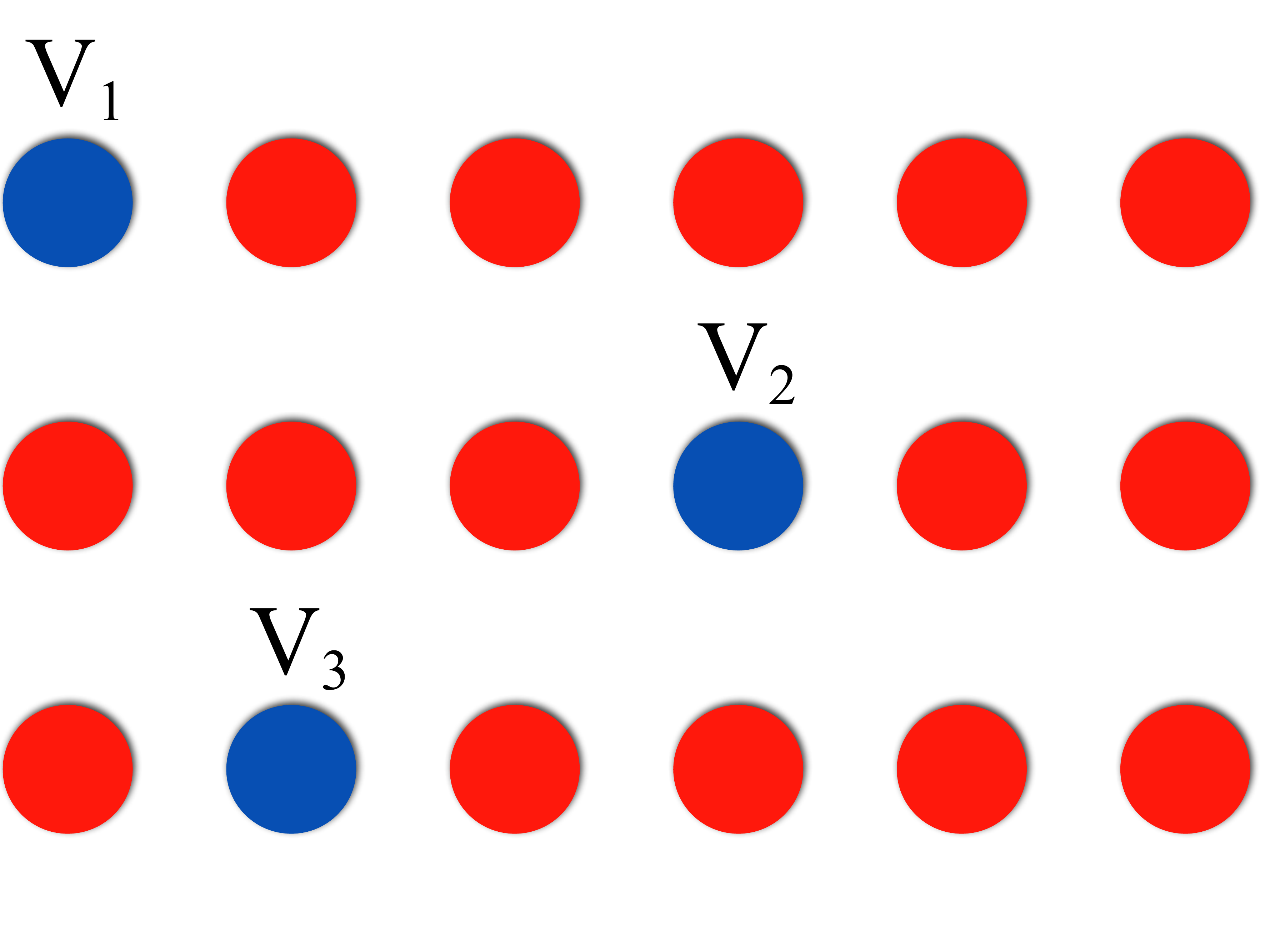}

(b)
\smallskip

\includegraphics[width=8.3cm]{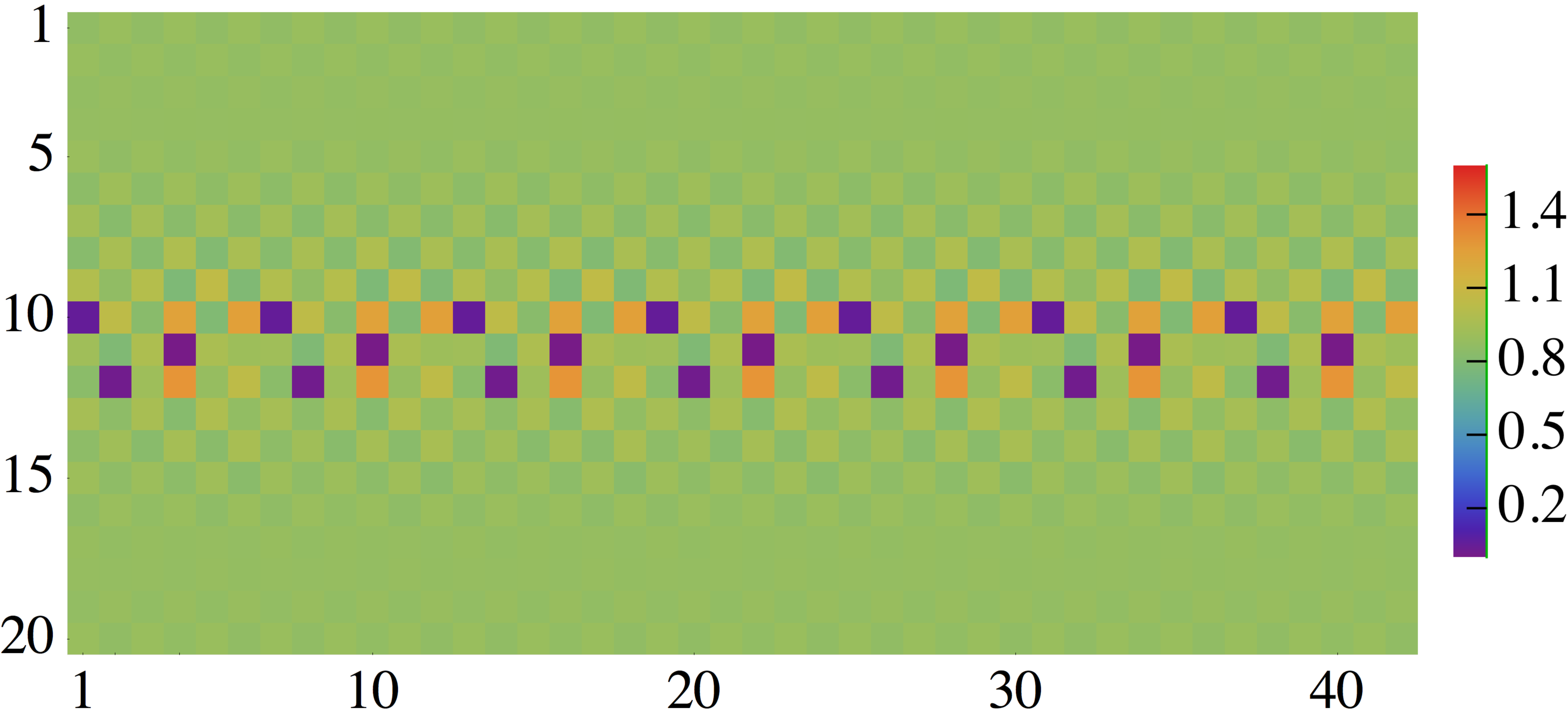} 
\end{center}
\caption{(Color online) GB with hopping amplitudes from Fig.~\ref{fig:GBhopping} and potential scatterers $V_{1,2,3}$.  (a) Scheme for the potential scatterers in a GB structural unit. (b) Electron occupation number $n_i$ for a (42$\times$20) site system. The potentials  are $V_1=10\,t$ in line~10, $V_2=20\,t$ in line~11, and $V_3=10\,t$ in line~12 according to the scheme of (a). The occupation is minimal on the sites (in blue) where the potentials are set. The control parameters are $U=2t$, $T=0.11\,t$, and the filling is$n=0.86$. The local magnetization is zero for this configuration and is therefore not displayed.}
\label{fig:V1V2V3}
\end{figure}

A different scenario can be generated with a specific choice of a scattering-potential profile. In Fig.~\ref{fig:V1V2V3} three potential scatterers have been introduced in the GB structural unit. This set-up results in a suppression of magnetism along the GB (not displayed in Fig.~\ref{fig:V1V2V3}). The origin for this suppression is tied to the distinct distribution of occupation numbers: the strong on-site potentials not only annihilate the magnetic moments at the respective sites of the scatterers but also induce a sizable increase in the electronic occupation of nearby sites (Fig.~\ref{fig:V1V2V3}b) well beyond single occupation; there the occupation number is close to 1.3. Consequently, in this set-up, the occupation numbers are either high or low at all the sites which carried the strong magnetic moment in the absence of on-site potentials.

It remains to be examined if the latter scenario with a suppression of magnetism is more realistic for the actual cuprate GBs than the scenario with robust magnetism. A more detailed analysis with data from electronic structure evaluations has to be implemented which, however, is not feasible at present. Certainly, the potential scattering is strong at sites where nearby oxygen atoms are missing. The oxygen vacancies also misalign the positions of the Cu sites which, in the majority of the cases, leads to a smaller hopping amplitude to nearest-neighbor sites (see Fig.~3 of Ref.~\onlinecite{Graser}). Those neighbor sites with a reduced hopping amplitude would probably form magnetic moments but a higher occupation of the sites could suppress the magnetic moment. Although such correlations between sites with strong potentials and bonds with reduced hopping amplitudes and sizable shifts in site occupation exist, it is not clear from the previous evaluations if magnetism is suppressed or rather assisted. This competition has to be explored in a prospective investigation. 

\begin{figure}[b]
\centering
\includegraphics*[width=4.2cm]{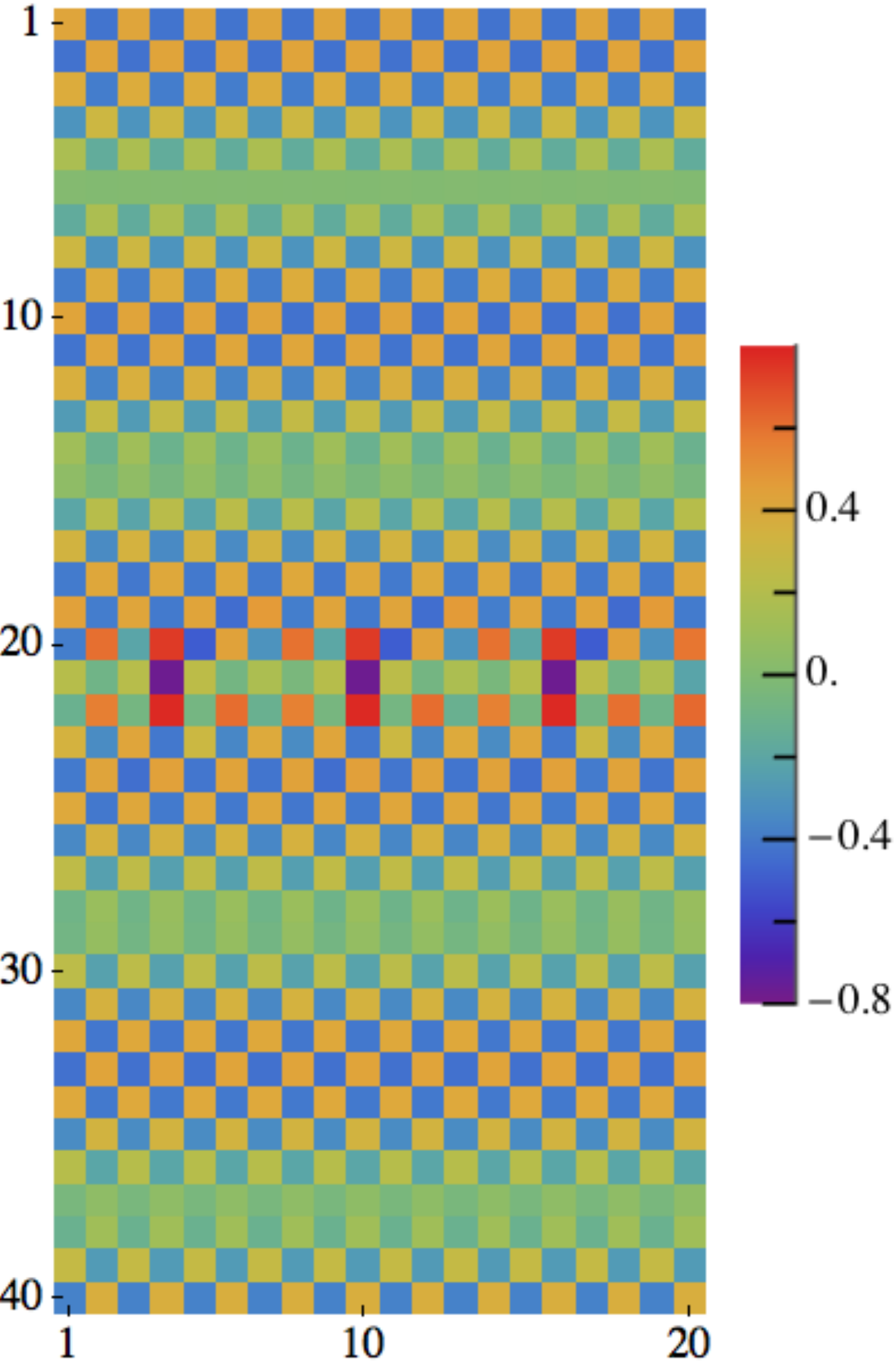}
\includegraphics*[width=4.072cm]{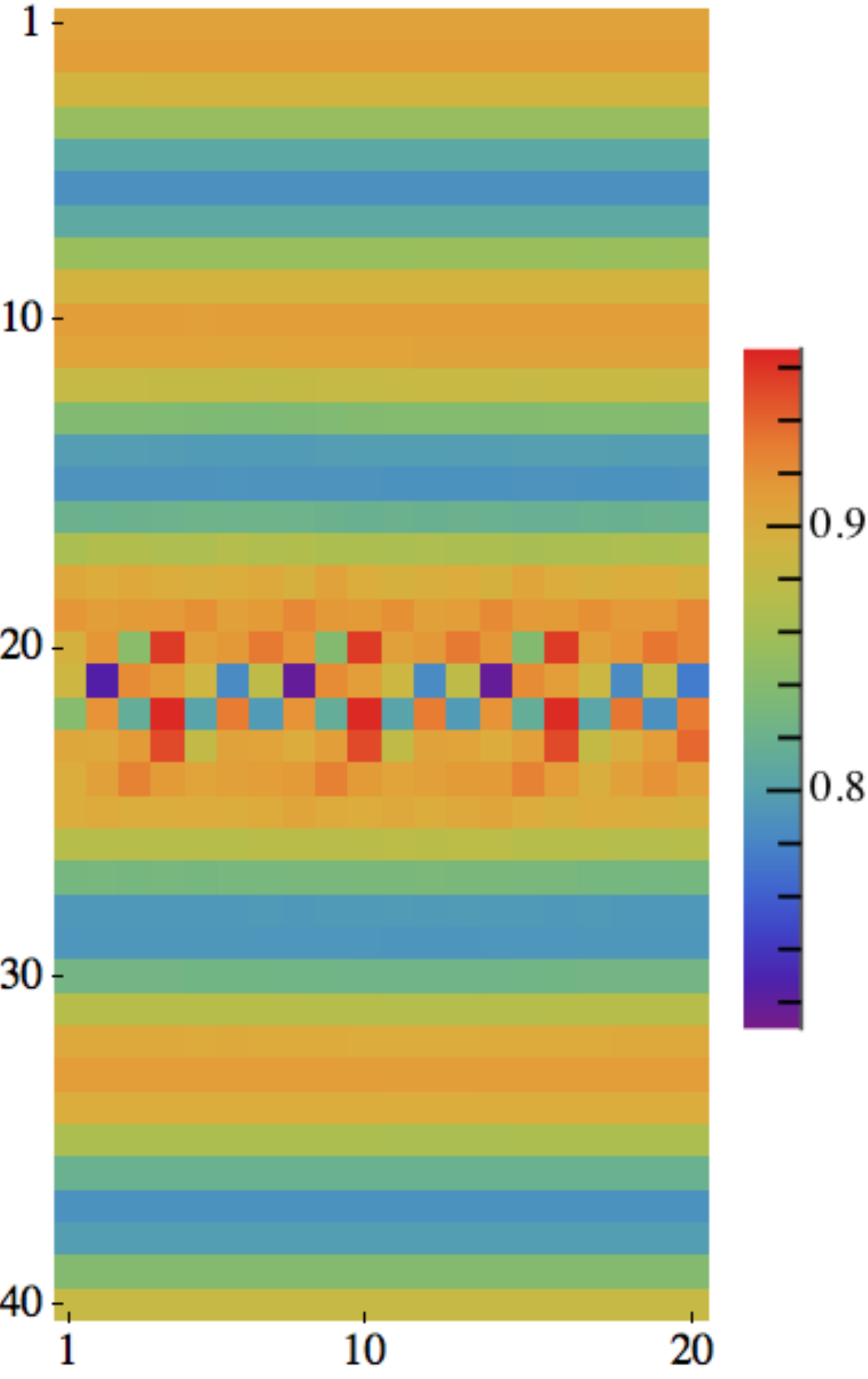}
\caption[]{(Color online) GB-induced stripes. Local magnetization $m_i$ (left panel) and electron occupation number $n_i$ (right panel) at the GB defined through Fig.~\ref{fig:GBhopping}  for $U=3t$, $T=0.08\,t$, and filling $n=0.86$.\label{fig:stripes}}
\end{figure}

Finally, we readdress the magnetic oscillations, i.e.\ magnetic stripes, which extend from the GB into the bulk where they decay after several periods. It is well established that models with built-in electronic correlations display stripe states~\cite{Zaanen,Tranquada,Kivelson,Raczkowski} in real space mean-field evaluations (see, for example, Refs.~\onlinecite{Schmid,Loder1,Loder2}). Here, the stripes are induced by the inhomogeneity of the GB but the chosen value of $U/t$ is not sufficiently large to support them in the bulk phase. The GB-induced stripes (Fig.~\ref{fig:stripes}) are separated by nonmagnetic lines with lower electron occupation, which constitute antiphase domain walls. The closer to half-filling the wider the stripes, in agreement with previous results for the bulk stripe phase. Moreover, stripes can be pinned or induced by impurities and line defects.~\cite{Andersen2,Schmid}

\section{Normal state transport at the GB}\label{sec:transport}

The resistance of a GB can be calculated either from the Landauer approach to mesoscopic conductors and barriers or from the Kubo formula for an inhomogeneous electronic system (see Sec.~5 in the book by Y.~Imry~\cite{Imry}\ for an introductory discussion, and Refs.~\onlinecite{Economou}, \onlinecite{Lee}, and \onlinecite{Stone} for the compatibility of the two approaches). It is not our intention to investigate the normal state transport at the GB in depth. In connection with the considered local moment formation we want to find the pattern of current densities at the GB and determine the GB resistance $R(T)$. For this purpose the evaluation of the Kubo linear response formula is best suited. Summation of the current densities over appropriate bonds in a line parallel to the GB will allow to identify the total current and consequently the resistance of the GB system. The resistance $R(T)$, as derived from the Kubo formula, is to be identified with that from a 2-terminal measurement in an experimental determination of $R(T)$.~\cite{Economou,Lee,Stone,Imry} However, with the large number of channels in the 2D setup, the contact resistance contributes little to $R(T)$, and the result of the Kubo formula evaluation may be approximately associated with the GB resistance. We like to mention that the criteria by Scalapino, White and Zhang~\cite{Scalapino}, which allow beautifully to distinguish between insulating, metallic and superconducting states through the zero-frequency limit of the current-current correlation function, do not apply here, as the GB breaks the translational invariance in all directions and one cannot obtain the long-wavelength ${\bf q\rightarrow 0}$ limit. The evaluation of the current-current response has to be performed in real space.

In linear response theory the non-local conductivity is calculated from the commutator of the paramagnetic part of
the current operator ${\bf j^{\rm p}}$ through

\begin{widetext}
\begin{align} 
\label{eq:conductivity}
\sigma_{\alpha\beta}({\bf r},{\bf r'},\omega)\!=\!
\frac{1}{i\omega}\biggl(\int_{-\infty}^\infty \!\! d(t-t')e^{i\omega(t-t')}(\frac{-i}{\hbar})
\times\theta(t-t'))
\langle[j^{\rm p}_{\alpha}({\bf r},t),j^{\rm p}_{\beta}({\bf r'},t')]\rangle
-\langle \sum_\sigma  t_{{\bf r, r+a_\alpha}} c^\dagger_{{\bf r}\sigma}c^{\phantom{\dagger}}_{{\bf r+a_\alpha}\sigma}+{\rm h.c.}\rangle
\,\delta_{\alpha,\beta}\delta_{{\bf r},{\bf r'}}\biggr).\nonumber
\end{align}
\end{widetext}
The paramagnetic component of the current operator is expressed by
\begin{equation}
j^{\rm p}_{\alpha}({\bf r})=\frac{i e}{2\hbar}c^\dagger_{{\bf r}\sigma}
(t_{{\bf r, r+a_\alpha}}c_{{\bf r+a_\alpha}\sigma}-t_{{\bf r, r-a_\alpha}}c_{{\bf r-a_\alpha}\sigma})\,+\,{\rm h.c.}
\end{equation}
where the vector ${\bf r+a_\alpha}$ is the position of the nearest neighbor site to ${\bf r}$ in the direction indicated by
the index $\alpha$.
With the unitary transformation onto fermionic operators $\gamma^\dagger_{m\sigma}$ and $\gamma_{m\sigma}$:
\begin{equation}
c^\dagger_{{\bf r}\sigma}= \sum_m u^\star_{m\sigma{\bf r}}\gamma^\dagger_{m\sigma},\quad 
c_{{\bf r}\sigma}= \sum_m u_{m\sigma{\bf r}}\gamma_{m\sigma}
\end{equation}
one diagonalizes the Hamiltonian~(\ref{eq:hamiltonian}). One finds for the non-local dc conductivity:

\begin{widetext}
\begin{align}\label{eq:dcconductivity}
\sigma^{\rm dc}_{\alpha\beta}({\bf r_1},{\bf r_2}) = \,&\frac{e^2 \pi}{4a^2\hbar}\,\lim_{\omega\to0}\,\sum_{m,n,\sigma} \frac{f(E_m)-f(E_n)}{\hbar\omega}\times\frak{D}(E_m - E_n +\hbar\omega, \Delta)\\ 
&\!\!\times\! \biggl[ \!
\bigl(t_{{\bf r_1,r_1-a_\alpha}}u^*_{m\sigma({\bf r_1-a_\alpha})}\!\!-t^*_{{\bf r_1,r_1+a_\alpha}}u^*_{m\sigma({\bf r_1+a_\alpha})}\bigr)u_{n\sigma{\bf r_1}}u^*_{n\sigma{\bf r_2}}\bigl(t_{{\bf r_2,r_2+a_\beta}}u_{m\sigma({\bf r_2+a_\beta})}\!\!-t^*_{{\bf r_2,r_2-a_\beta}}u_{m\sigma({\bf r_2-a_\beta})}\bigr)\nonumber\\
&+\bigl(t_{{\bf r_1,r_1+a_\alpha}}u_{m\sigma({\bf r_1+a_\alpha})}\!\!-t_{{\bf r_1,r_1-a_\alpha}}^*u_{m\sigma({\bf r_1-a_\alpha})}\bigr)u^*_{n\sigma{\bf r_1}}u_{n\sigma{\bf r_2}}\bigl(t_{{\bf r_2,r_2-a_\beta}}u^*_{m\sigma({\bf r_2-a_\beta})}\!\!-t^*_{{\bf r_2,r_2+a_\beta}}u^*_{m\sigma({\bf r_2+a_\beta})}\bigr)\nonumber\\
&+\bigl(t_{{\bf r_1,r_1-a_\alpha}}u_{m\sigma({\bf r_1 -a_\alpha})}^*\!\!-t_{{\bf r_1,r_1+a_\alpha}}^*u^*_{m\sigma({\bf r_1 +a_\alpha})}\bigr)u_{n\sigma{\bf r_1}}u_{m\sigma{\bf r_2}}\bigl(t_{{\bf r_2-a_\beta,r_2}}u^*_{n\sigma({\bf r_2-a_\beta})}\!\!-t_{{\bf r_2,r_2+a_\beta}}^*u_{n\sigma({\bf r_2+a_\beta})}^*\bigr)\nonumber\\
&+\bigl(t_{{\bf r_1,r_1+a_\alpha}}u_{m\sigma({\bf r_1+a_\alpha})}\!\!-t_{{\bf r_1,r_1-a_\alpha}}^*u_{m\sigma({\bf r_1-a_\alpha})}\bigr)u^*_{n\sigma{\bf r_1}}u^*_{m\sigma{\bf r_2}}\bigl(t_{{\bf r_2,r_2+a_\beta}}u_{n\sigma({\bf r_2+a_\beta})}\!\!-t^*_{{\bf r_2,r_2-a_\beta}}u_{n\sigma({\bf r_2-a_\beta})}\bigr)\biggr]\nonumber
\end{align}.
\end{widetext}
In the absence of a magnetic field, the $t_{{\bf r_i,r_i\pm a_\alpha}}$ are real and the coefficients $u_{m\sigma{\bf r_i}}$ may be chosen real.

The dissipative part of the response function~(\ref{eq:dcconductivity}) is controlled by the Dirac $\delta$-function, i.e., $\frak{D}(E_m - E_n +\hbar\omega, \eta)$ is in fact $\delta(E_m - E_n +\hbar\omega)$ for the system defined by the Hamiltonian~(\ref{eq:MFhamiltonian}) with respective eigenvalues $E_m$. As the system is finite, the spectral function is composed of $\delta$-functions. The true GB system is however coupled to a bath with a continuum of excitations. This bath may be provided by phonons or by the leads. The standard scheme to allow for dissipation of a finite system coupled to a bath is to replace the $\delta$-functions by Gau{\ss} functions $\frak{D}(E_m - E_n +\hbar\omega, \eta)$ of width $\eta$ so that the spectrum becomes continuous. This has been discussed extensively in the literature on mesoscopic electronic systems (see, for example, Ref.~\onlinecite{Imry}, and references therein). We consider the case where $\eta$ is larger than the distance between adjacent energy levels.

\begin{figure}[t]
\begin{center}
(a)
\includegraphics[width=8.9cm]{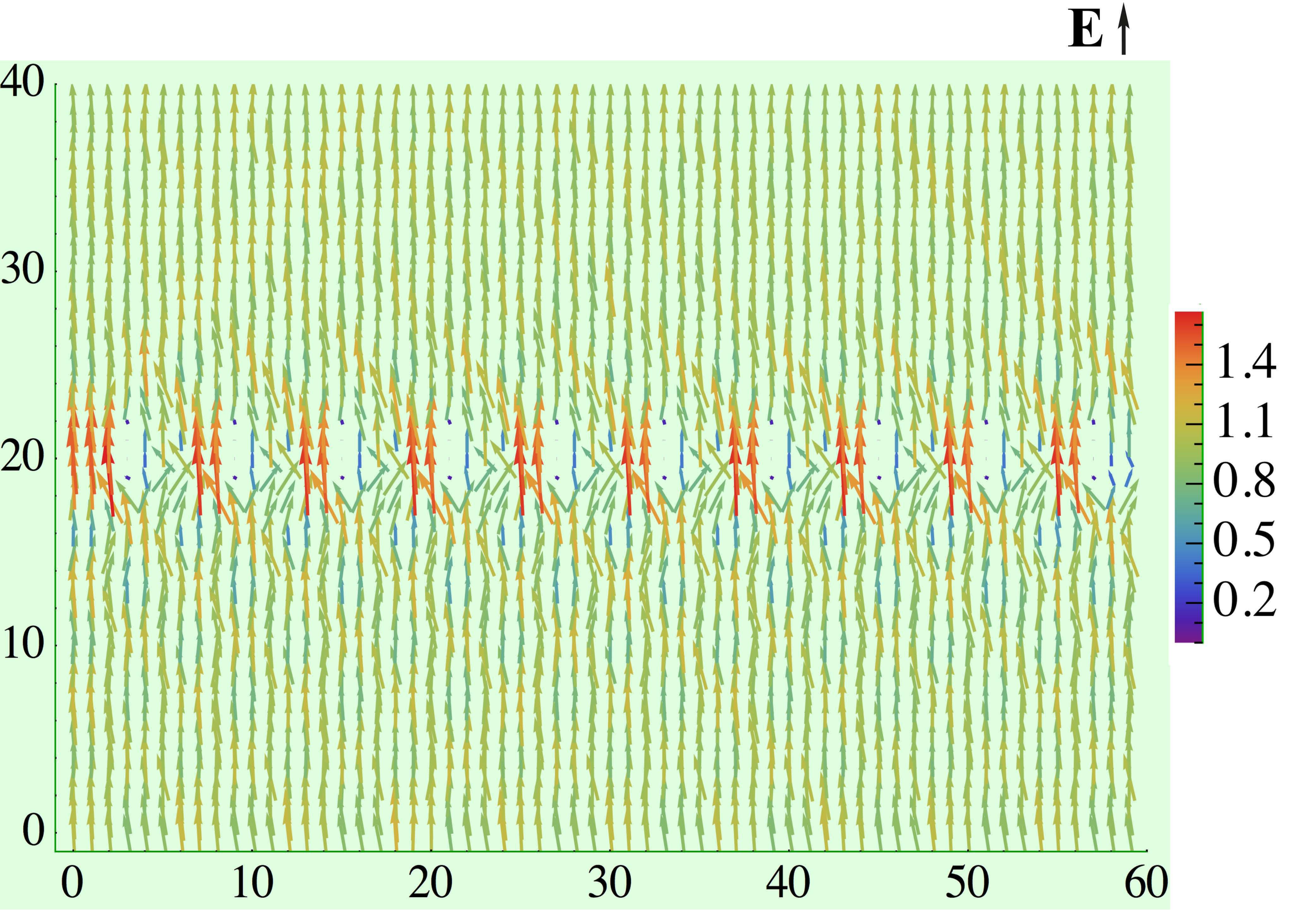}
\smallskip

(b)
\smallskip

\includegraphics[width=8.9cm]{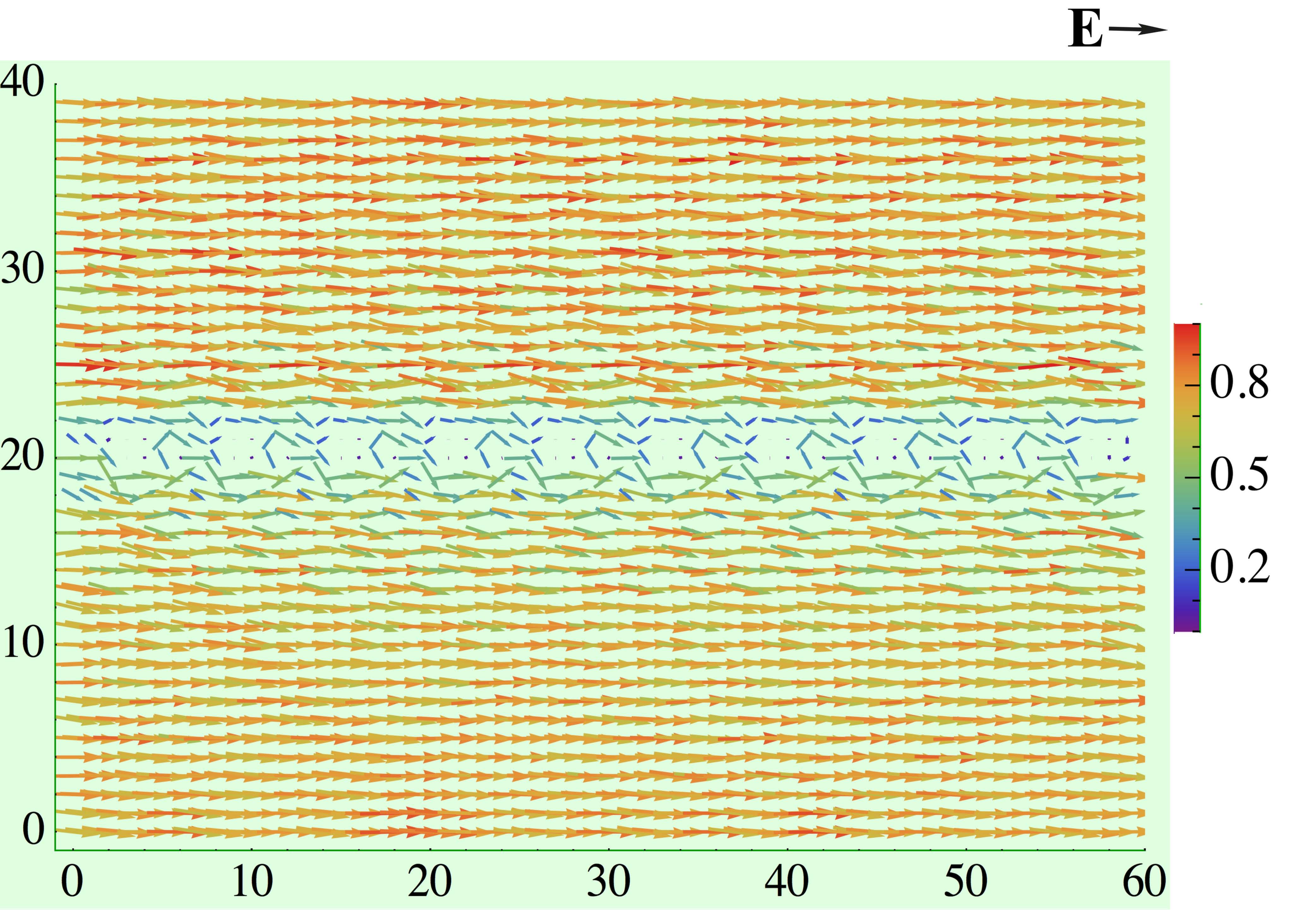}
\end{center}
\caption{(Color online) Current density pattern for a GB with hopping amplitudes from Fig.~\ref{fig:GBhopping}. The external field is perpendicular to the GB in the panel (a) and parallel to the GB in panel (b). The local current densities (in a.u.)\  are determined 
for $U=2t$, $T=0.05\,t$ and filling $n=0.86$ from Eqs.~(\ref{eq:dcconductivity}) and (\ref{eq:current}).}
\label{fig:CurrentDensity}
\end{figure}

With the determination of $\sigma^{\rm dc}_{\alpha\beta}({\bf r},{\bf r'})$ through the eigenvectors and eigenvalues of Hamiltonian~(\ref{eq:MFhamiltonian}) 
one may evaluate the current density ${\bf j}({\bf r})$ at any point in the system for given electric field ${\bf E}({\bf r'})$:
\begin{equation}
\label{eq:current}
j_{\alpha}({\bf r})=\sum_{\bf r'}\sum_{\beta}\sigma^{\rm dc}_{\alpha\beta}({\bf r},{\bf r'}) E_\beta({\bf r'})
\end{equation}
Here we assume a constant field ${\bf E_0}$ across the system. This evaluation neglects charge inhomogeneities and the corresponding screening. In order to cope with these effects one would have to include non-local Coulomb interaction terms which however is beyond the present assessment based on the Hubbard model. The values of screening lengths in the cuprates are not precisely known but near optimal doping they are expected to be of the order of a lattice spacing or less (cf.\ Ref.~\onlinecite{Graser}). Correspondingly, we estimate that the corrections due to non-local Coulomb interactions do not change our predictions qualitatively. 

The pattern of local current densities is displayed in Fig.~\ref{fig:CurrentDensity}. The apparent feature is the formation of conducting channels. This property is anticipated because the hopping amplitudes are small on the bonds in the area between the channels. However, it is important to realize that the magnetic moments are formed in this area. Correspondingly, we have the scheme that reduced bond kinetic energies within the GB allow for the formation of local magnetic moments if the local electron occupations are not far from one and if on-site Coulomb repulsion is not too small (larger than the bond kinetic energies). These regions with reduced hopping amplitudes block the current through the GB and give rise to current channels with a width of interatomic Cu distances. 
The current pattern for an electric field parallel to the GB (Fig.~\ref{fig:CurrentDensity}b) is consistent with that for fields perpendicular to the GB (Fig.~\ref{fig:CurrentDensity}a). In both cases one observes interference patterns from the periodically repeated conducting channels. These patterns extend well into the bulk: the interference produces deviations from the uniform bulk current density of 25\% at a distance of the order of the GB width (measured from the GB edge). The GB is not mirror symmetric, and therefore one observes slight deviations in the current patterns above and below the GB in Fig.~\ref{fig:CurrentDensity}.

\begin{figure}[b]
\hspace*{-0.4cm}\includegraphics[width=9.5cm]{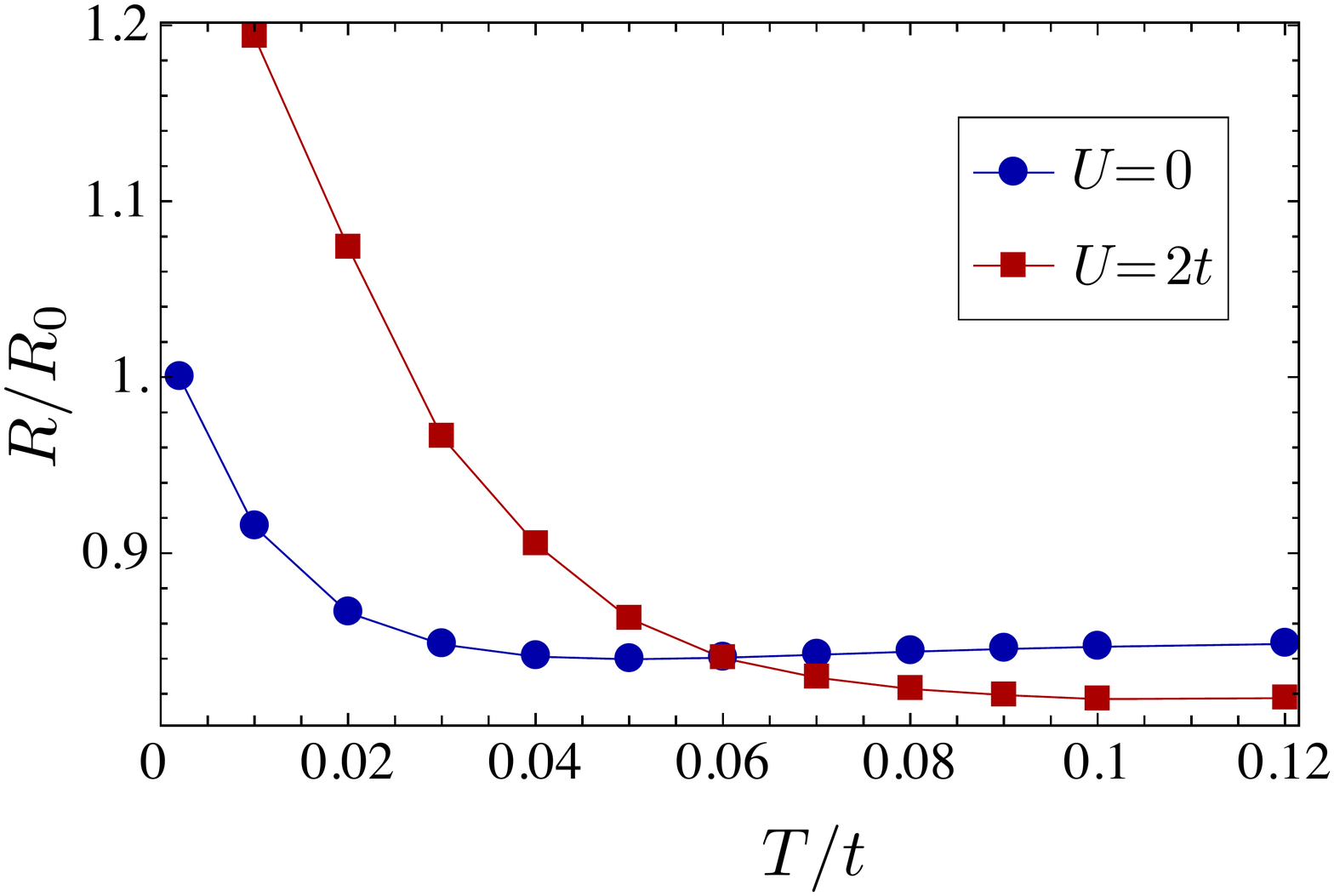}
\caption[]{(Color online) Temperature dependence of the GB resistance for $U=0$ (blue dots) and $U=2t$ (red squares) at $n=0.86$ for a (60$\times$40) site system. The resistance is normalized with respect to its value at the lowest evaluated temperature for $U=0$.
\label{fig:Resistance}}
\end{figure}

The resistance of a GB is controlled by a number of elastic and inelastic scattering processes, and not all of them are included in our evaluation---certainly, scattering on local phonon modes, on orbital or charge transfer excitations, and Kondo screening are not included. Here, we focus on the physics covered in our modelling, i.e., magnetic moment formation and the build-up of conducting channels and barriers at a GB with disorder in the hopping amplitudes. Again we take the distribution of hopping matrix elements assigned to the GB bonds in Fig.~\ref{fig:GBhopping}. The electric current $I$ through the GB system is identified from $I=\sum_{i \in L}j_{\perp}({\bf r_i})\cdot a$ where we sum over current density components in the direction perpendicular to the GB along a line $L$  parallel to the GB (in the bulk area). In fact, the current density components parallel to the GB sum up to zero for this situation with an ${\bf E}$ perpendicular to the GB. The lattice constant in the bulk is $a$. The conductance results from the relation $I=G\cdot V$ where the voltage drop $V$ across the system is determined by ${\bf E}$. The resistance $R(T)$ is $1/G(T)$ and we evaluated $R(T)$ in the temperature range $0.01\,t \leq T \leq 0.22\,t$ with finite GB magnetization. The magnetic moments are lost for temperatures above approximately $T_c\sim 0.22\,t$. 

The resistance shows a non-monotonous temperature behavior: it decreases slightly for decreasing temperatures (at the high-temperature side) and then increases towards low temperatures. This behavior is observed for both $U=0$ and $U=2t$, however, the resistance minimum is shifted to higher temperatures for the $U=2t$ and the increase on the low temperature side is significantly more pronounced for the GB with magnetic moments (see Fig.~\ref{fig:Resistance}). 

\begin{figure}[t]
\hspace*{-0.3cm}\includegraphics*[width=8.9cm]{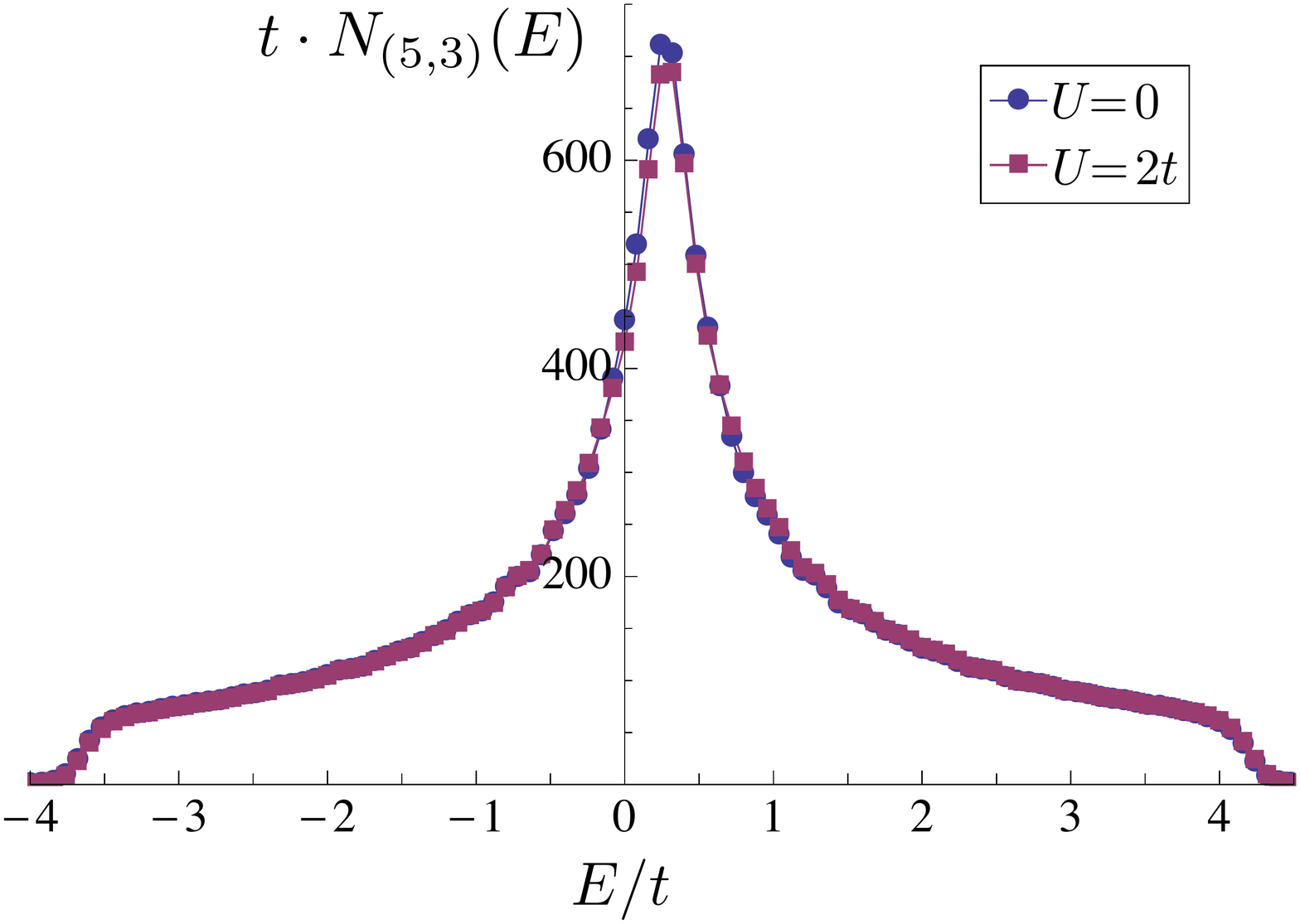}
\caption[]{(Color online) LDOS $N_i(E)$ for a site far from the GB. The site coordinate $i$ is at (5,3) in a $(60\times40)$ site system. The blue data points are calculated for $U=0$, the red for $U=2t$. The chemical potential is at zero energy and $T=0.02\,t$.\label{fig:LDOS}}
\end{figure}

The correlation-independent increase of $R(T)$ at the high temperature side is expected for this system and is related to the Fermi function factors in the expression for $\sigma^{\rm dc}(T)$. The two curves for $U=0$ and $U=2t$ converge for temperatures above the magnetic transition. This temperature dependence at the high temperature side may be masked by that of inelastic scattering processes in real systems. 

The increase of $R(T)$ for temperatures well below $0.1\,t$ is better suited to characterize the investigated GB system. The local current at the GB is controlled by three distinct physical factors in $\sigma^{\rm dc}(T)$ of Eq.~(\ref{eq:dcconductivity}): the density of states, the correlation of electron momenta, which transit through the GB barrier, and the relaxation time of inelastic scattering processes. 

The latter is implemented in $\frak{D}(E_m - E_n +\hbar\omega, \eta)$ through a finite broadening $\eta$.  
We assume a temperature independent broadening in this work and focus on the impact of a magnetic GB state on the density of states (DOS) and electronic momenta. 

\begin{figure}[t]
\vspace*{-0.25cm}
\hspace*{-0.3cm}
\includegraphics[width=9.5cm]{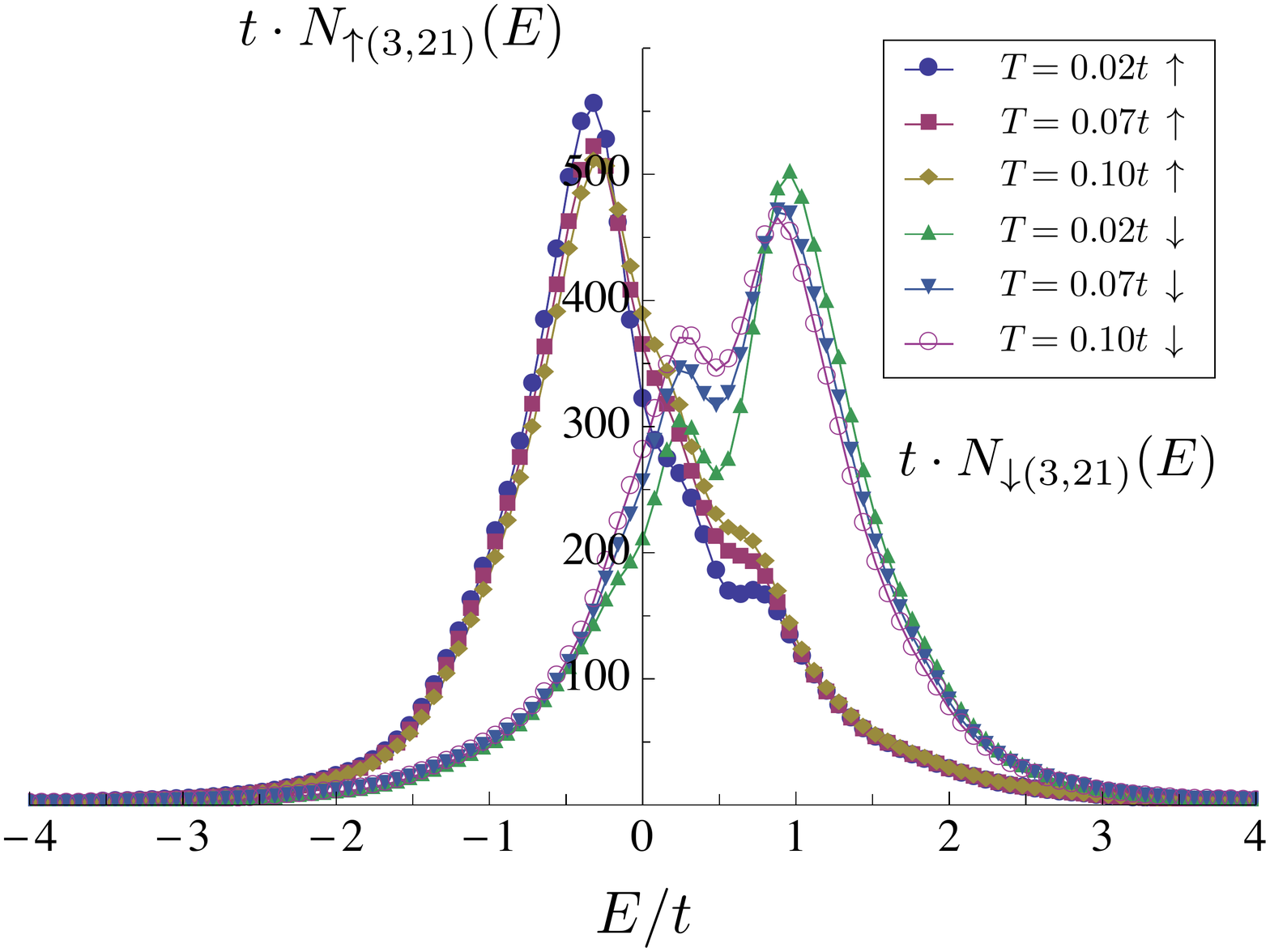}
\caption[]{(Color online) Spin-resolved LDOS for a site at the GB that is blocked by a local moment (between two conducting channels). The site coordinate is at (3,21) in a $(60\times40)$ site system. The curves, which are peaked below the chemical potential (at zero energy), present the LDOS for the up-spin direction with blue points at $T=0.02\,t$, red at $T=0.07\,t$, and beige at $T=0.1\,t$. The curves with the major part of their weight above the chemical potential are the corresponding LDOS data points for down-spin direction.\label{fig:SRLDOS}}
\end{figure}

The local density of states (LDOS) for a site $i$ is calculated from $N_i(E)=\sum_{m,\sigma} u^\star_{i,m,\sigma} u_{i,m,\sigma} \delta(E-E_m)$ and one obtains the DOS from $N(E)=\sum_{i} N_i(E)$. Whereas the shape of the LDOS at sites distant from the GB is a smoothed-out DOS of an infinite square lattice (see Fig.~\ref{fig:LDOS}), the LDOS at GB sites deviates significantly from the bulk DOS. Most prominently, for sufficiently large values of $U$, the GB LDOS develops a dip above the Fermi energy for temperatures below the transition to a magnetic GB. This pseudo gap behaviour of the LDOS at sites within the GB is expected, as it reflects the formation of a magnetic state: a site with a magnetic moment has a high LDOS close to the Fermi edge for the corresponding spin direction whereas the opposite spin direction belongs to a high energy state above the pseudo gap (see the spin-resolved LDOS in Fig.~\ref{fig:SRLDOS}). 

Irrespective of the detailed dependence of $N_i(E)$ on site $i$, the DOS $N(E)$ times the static current-current correlation factor in Eq.~(\ref{eq:dcconductivity}) is a convex bended function at the Fermi energy for $U=0$. This function is not $T$-dependent for $U=0$ but the convex bended function, when multiplied by the derivative of the Fermi function, produces an increase in $\sum_i\sigma^{\rm dc}_{\alpha\beta}({\bf r_i},{\bf r_j})$ of Eq.~(\ref{eq:dcconductivity}) with increasing temperature. This observation explains the temperature dependence of $\sigma^{\rm dc}(T)$ for $U=0$---and evidently the low temperature dependence of $R(T)$ for $U=0$. We emphasize that this  effect is rather small and may depend on the detailed DOS and the proximity of the chemical potential to a van Hove singularity. 

\begin{figure}[b]
(a)\hfill\hfill

\includegraphics*[width=9.2cm]{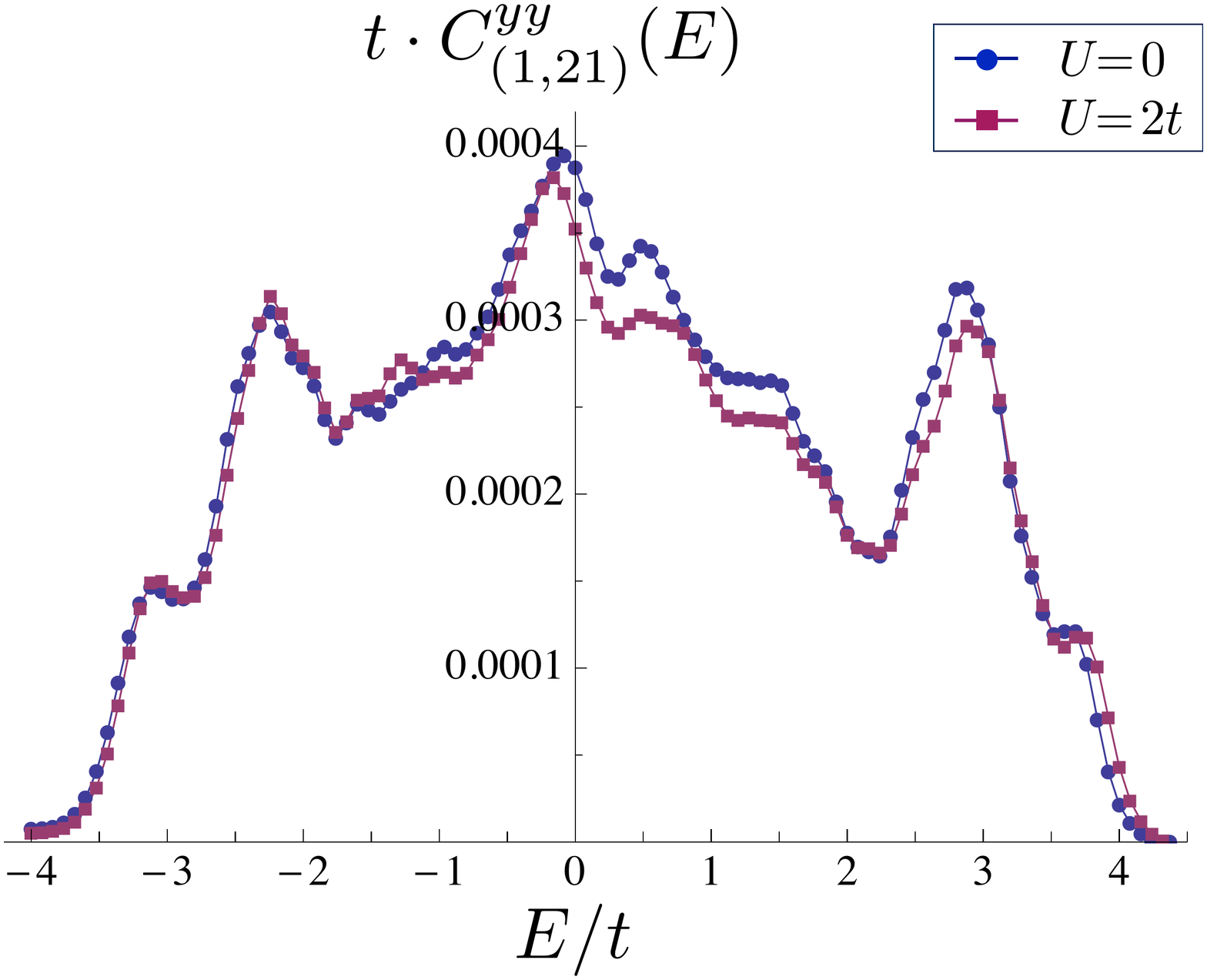}

\smallskip

(b)\hfill\hfill

\includegraphics*[width=9.1cm]{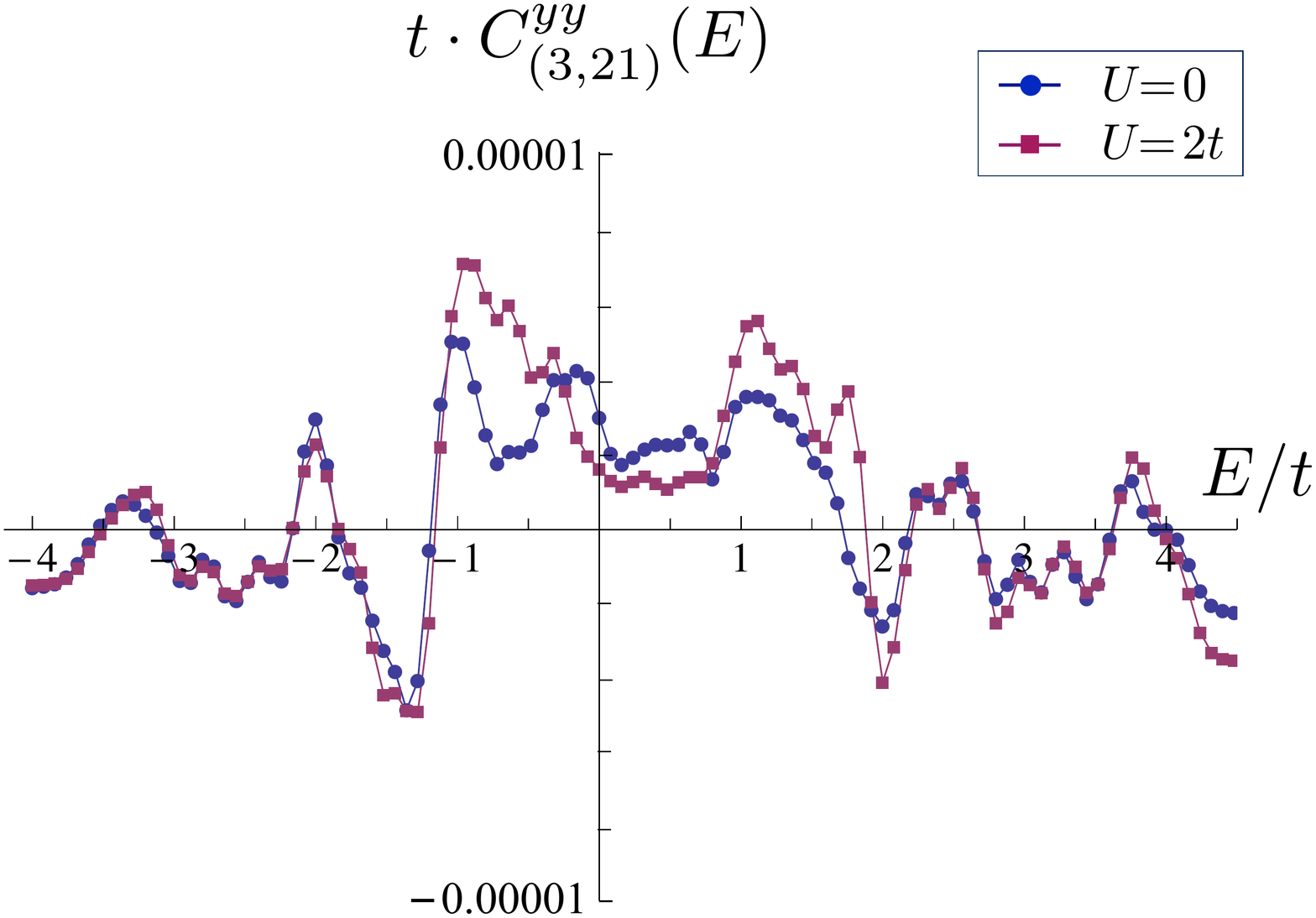}
\caption[]{(Color online) Local density of current correlations $C^{yy}_i(E)$ of Eq.~(\ref{eq:LDOC}). The $y$-direction is perpendicular to the GB. The upper panel displays $C^{yy}_i(E)$ for a site $i=(1,21)$ within a conducting channel.  The lower panel shows $C^{yy}_i(E)$ for a blocked site (3,21) in between two conducting channels of the GB. The temperature is $T=0.02\,t$, the blue data points are at $U=0$ and the red points at $U=2t$.\label{fig:LDOC}}
\end{figure}

\begin{figure}[b]
(a)\hfill\hfill

\includegraphics*[width=9.3cm]{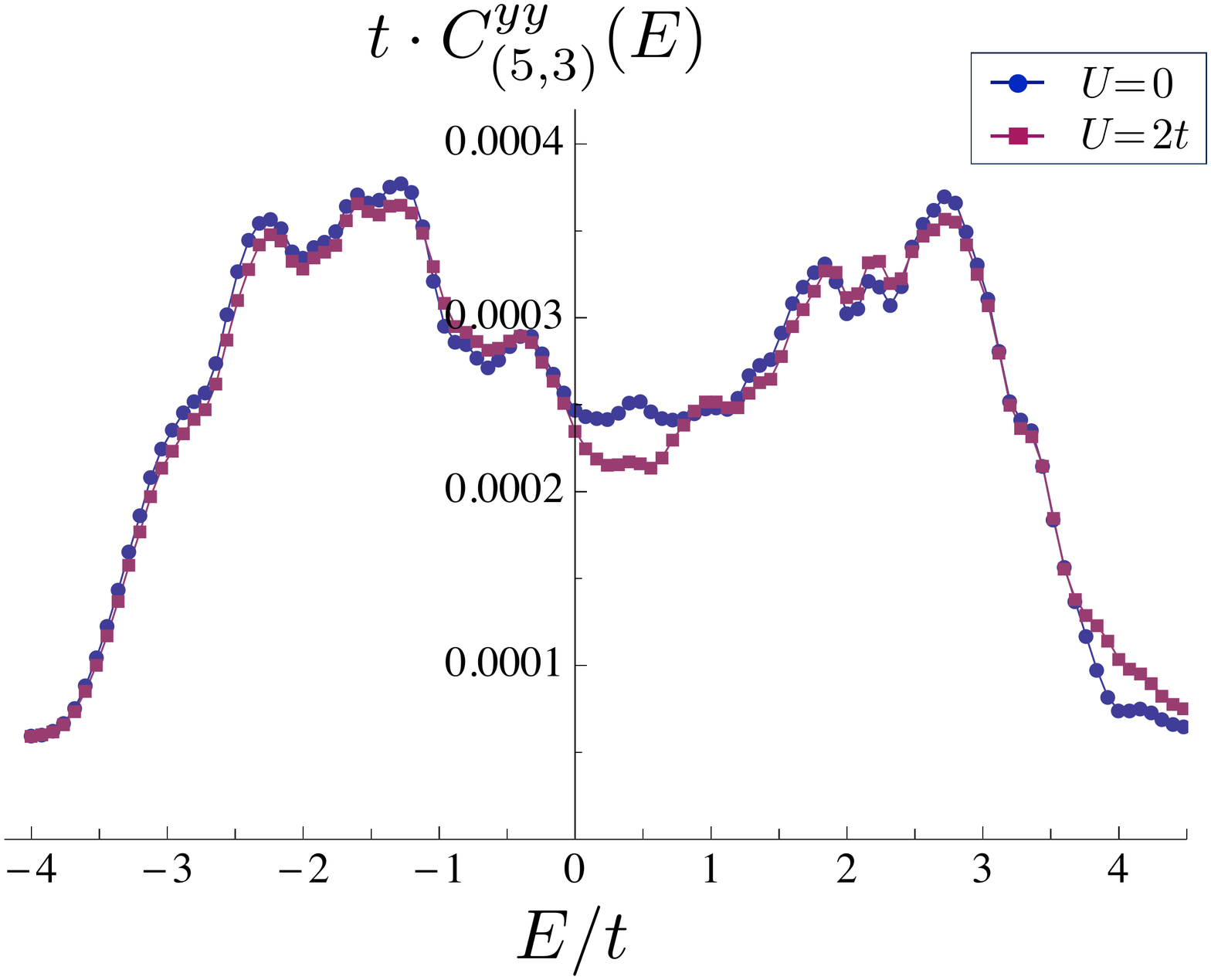}
\smallskip

(b)\hfill\hfill

\includegraphics[width=9.3cm]{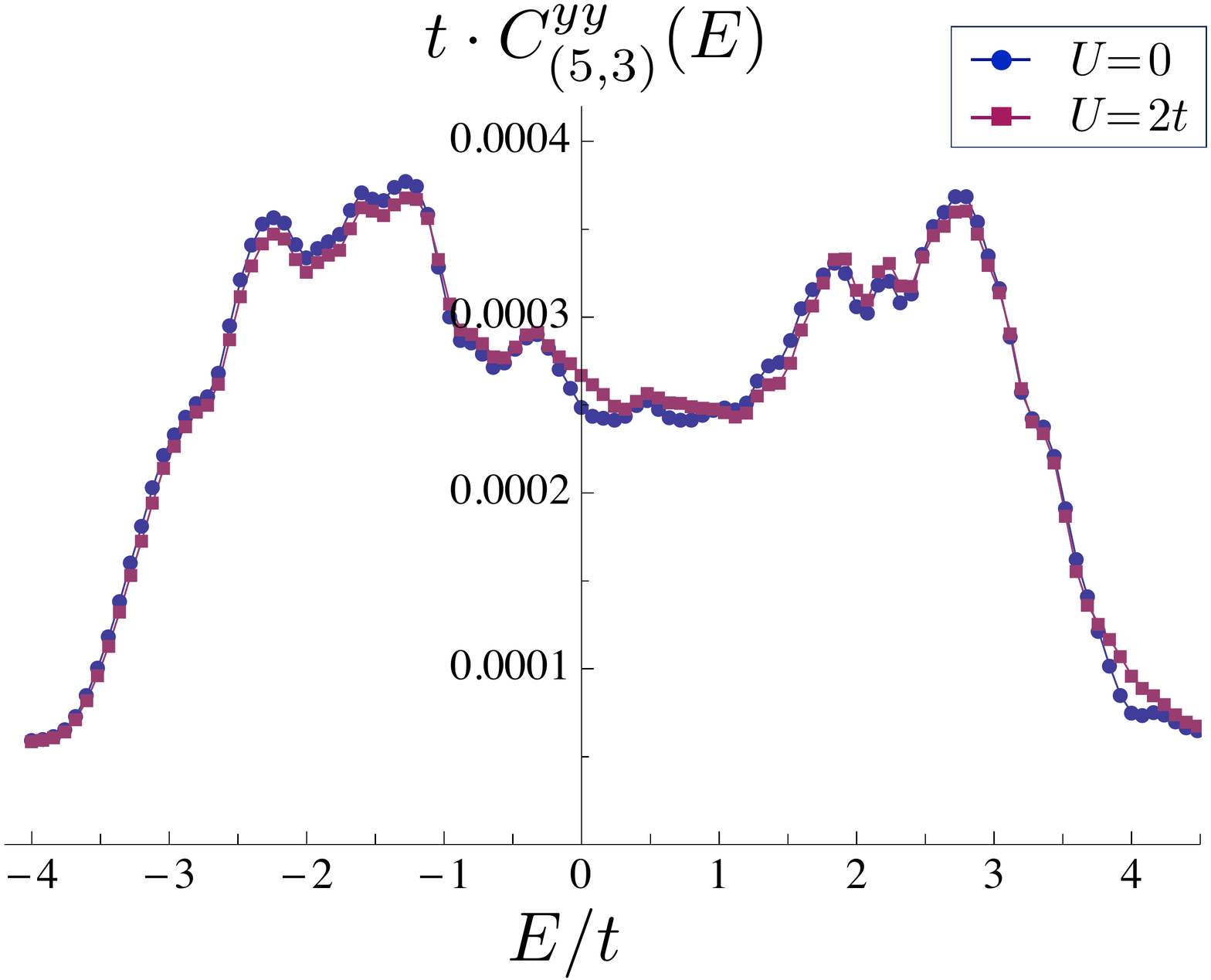} \hfill
\caption[]{(Color online) Local density of correlations $C^{yy}_i(E)$ of Eq.~(\ref{eq:LDOC}) for a site~$i=$~(5,3) far from the GB. The temperature is $T=0.02\,t$ for panel (a) and $T=0.12\,t$ for panel (b), respectively; the blue points were calculated at $U=0$ and the red points at $U=2t$.\label{fig:LDOC_bulk}}
\end{figure}

For finite on-site Coulomb interaction $U=2t$, a different mechanism causes the remarkably stronger increase of $R(T)$ in the range of the smallest temperatures at which $R(T)$ was evaluated (Fig.~\ref{fig:Resistance}).
The temperature dependence of $\sigma^{\rm dc}(T)$ of Eq.~(\ref{eq:dcconductivity})  and, consequently, of $R(T)$ is also controlled by the magnitude of the current-density correlations.  In Eq.~(\ref{eq:dcconductivity}), the current-density correlations are expressed by the quartic terms in the difference of neighbouring state eigenvectors $u_{m{\bf r_i \pm a_\alpha}}$ times the corresponding hopping matrix elements $t_{{\bf r_i, r_i+a_\alpha}}$ at two distinct sites ${\bf r_1}$ and ${\bf r_2}$  within the entire system. For the purpose to estimate this contribution we introduce a correlation function $C^{\alpha\beta}_i$ which is a local density of current correlations and which depends on the direction ${\bf a_\alpha}$ of the local current and  ${\bf a_\beta}$ of the applied electric field (the current correlation factor of Eq.~(\ref{eq:dcconductivity}) is taken real as we do not consider an external magnetic field here): 
\begin{align}  
\label{eq:LDOC}
C^{\alpha\beta}_i &(E)= \sum_{m,n,\sigma}\;\sum_{{\bf r_j}}\; 
\delta(E-E_n)\;\times\\
\biggl[&\; u_{m\sigma{\bf r_1}}\left(t_{{\bf r_1,r_1-a_\alpha}}u_{n\sigma({\bf r_1-a_\alpha})}\!\!-t_{{\bf r_1,r_1+a_\alpha}}u_{n\sigma({\bf r_1+a_\alpha})}\right)\nonumber\\
&\times\left(t_{{\bf r_2-a_\beta,r_2}}u_{m\sigma({\bf r_2-a_\beta})}\!\!-t_{{\bf r_2,r_2+a_\beta}}u_{m\sigma({\bf r_2+a_\beta})}\right)u_{n\sigma{\bf r_2}}{\phantom{\int}}\nonumber\\
&\!\!\!\!\!+\! u_{n\sigma{\bf r_1}}\left(t_{{\bf r_1,r_1+a_\alpha}}u_{m\sigma({\bf r_1+a_\alpha})}\!\!-t_{{\bf r_1,r_1-a_\alpha}}u_{m\sigma({\bf r_1-a_\alpha})}\right){\phantom{\int}}\nonumber\\
&\times\left(t_{{\bf r_2,r_2-a_\beta}}u_{m\sigma({\bf r_2-a_\beta})}\!\!-t_{{\bf r_2,r_2+a_\beta}}u_{m\sigma({\bf r_2+a_\beta})}\right)u_{n\sigma{\bf r_2}}
       \biggr]\nonumber\\ \nonumber
\end{align}

Evidently, a site which is blocked by a local moment will contribute less to the local current density than sites which form a conducting channel through the GB barrier.  In Figs.~\ref{fig:LDOC}(a) and \ref{fig:LDOC}(b) we display the energy resolved current-correlation factor $C^{yy}_i$ at various temperatures for sites $i$ in a conducting channel and blocked sites, respectively. The direction $y$ is perpendicular to the GB. Clearly, $C^{yy}_i$ is reduced in the GB magnetic state.
Also at sites $i$ far from the GB, $C^{yy}_i$ is suppressed in the GB magnetic phase (see Fig.~\ref{fig:LDOC_bulk}): the transformation vectors $u_{m{\bf r_i}}$ at site $\bf r_i$ depend on state $m$ of the system and therefore carry the information of the GB state even though $\bf r_i$ might be chosen far from the GB. This suppression of the current correlations, expressed by $C^{yy}_i$, is the dominant mechanism for the decrease of $\sigma^{\rm dc}(T)$ with decreasing temperature in the low-$T$ regime in our GB model. In fact, Fig.~\ref{fig:LDOC}(a) clearly shows for $T=0.02\,t$ that the current-density correlations $C^{yy}_i$ close to the Fermi energy are smaller for $U=2t$ as compared to $U=0$. 

At more elevated temperatures (see Fig.~\ref{fig:LDOC}(b) with $T=0.12\,t$) the current correlations $C^{yy}_i$ may even be slightly larger for finite $U$ close to the Fermi energy which is reflected in the lower value of $R(T)$ for $U=2t$ for this temperature range (see Fig.~\ref{fig:Resistance}). Eventually, for $T$ at the magnetic transition, the resistance curves for $U=2t$ and $U=0$ merge (not displayed in Fig.~\ref{fig:Resistance}), as $C^{yy}_i$ is temperature independent for the disordered state.

The temperature, at which the GB becomes magnetic, provides a scale compatible with the temperature at the minimum of $R(T)$ for intermediate values of $U$\/. However, the exact turning point depends on the details of the set-up of the GB. In particular, this estimate is valid for GBs which are formed by a reduction of the bond kinetic energies at the GB. A special distribution of the GB scattering potentials may have a considerable impact on this temperature.

Wei Chen {\it et al.}~\cite{WeiChen} attributed upturns in the resistivity of underdoped cuprates at low temperature to randomly distributed magnetic droplets. It is the enlarged cross section due to the formation of local magnetic moments which generates these upturns in their modelling. The underlying physics appears to be similar to what we find for the GBs.
\section{Conclusions}
Grain boundaries of correlated electron systems, such as those investigated in layered oxide (high-$T_c$)  compounds, not only pose a challenge for keeping the GB-related reduction of the electrical current minimal but they are also of fundamental interest in connection with inhomogeneous heterostructures and their correlation controlled properties. In this article we explored the normal conducting state of a GB system artificially engineered  by an inhomogeneous two-dimensional Hubbard model in order to pursue a couple of basic but intricate issues: when can one expect a formation of local magnetic moments at the GB, and do they affect the transport properties of the GBs in a characteristic manner? In particular, does the formation of magnetic moments allow to interpret the observed (linear) increase of the GB resistance?

It does not come as a surprise that an inhomogeneous Hubbard model with strong on-site interaction $U$ generates local moments at sites which are least coupled to their surroundings. This has already been analyzed in correlated disordered systems, such as in heavily doped Si:P systems.~\cite{Lakner,Langenfeld} 

However, it has not been evident that a sufficiently strong variance $\langle\Delta t\rangle$ of the bond kinetic energies at a quasi one-dimensional GB structure induces a transition from a non-magnetic state to a magnetic GB state at a critical value of the variance $\langle\Delta t\rangle_c$. The mean-field evaluation may overestimate the transition temperature to the magnetic GB state but we found a transition already for a moderate value of $U=2t$ (where $t$ is the bulk value of the hopping amplitude) for $\langle\Delta t\rangle_c \simeq 0.5\, t$. 

The formation of local moments also depends on the distribution of site potentials $V_i$. It is apparent that a sizeable $V_i$ (with $|V_i| \gg t,U$) suppresses local moments as the site occupation is either considerably smaller or larger than 1 for positive and negative potentials, respectively. Nevertheless, we also identified site-potential profiles that assist the formation of magnetic GB states.
The site-potential related inhomogeneity can reduce the bond kinetic energies for specific profiles and allow stronger magnetic moments at nearby sites. Future work, implementing a realistic non-stoichiometric GB composition, has to settle if either scheme applies and, consequently, GB magnetism is suppressed or enhanced  in the high-$T_c$ cuprates. 

The diagonalization of the GB system reveals that magnetism is not exclusively local in its appearance. The magnetic pattern reaches out into the vicinity of the GB on the scale of a few lattice constants. Moreover, the GB induces magnetic stripes in its proximity with a magnitude decaying with distance from the GB. The decay length of this phenomenon depends on $U$ which signals that beyond a critical value of $U/t>3$ the bulk develops the much investigated stripe state. The nonmagnetic lines in between the magnetic stripes exhibit lower electron occupation and form antiphase domain walls which is consistent with previous findings.

A distribution of hopping amplitudes and local scatterings potentials at the GB produces conducting channels if the `effective barrier' is not so strong as to block the current and allow only for tunnelling processes. For the GB profiles, which are considered in this work, we observe a distinct pattern of the current density at the GB. The 3-site wide channels also carry a current density when the electric field is applied in the direction parallel to the GB. The bulk current density is recovered only within one to two units of the GB width which is approximately the same length scale for an effective GB width as deduced from the magnetic pattern.

The most striking result of the transport properties is the increase of the resistance for decreasing temperature in the regime $T \lesssim 0.1\,t$.  For finite on-site repulsion $U$ we identify a strong enhancement of $R(T)$---a 50\% increase at $T/t=0.01$ from its minimal value at $T/t\simeq 0.1$. We relate this result to the formation of local magnetic moments at the GB. The prominent transport feature is the suppression of the current correlations in the magnetic GB state. This suppression controls the low-temperature resistance $R(T)$. 

We find a linear increase of $R(T)$ for the smallest temperature range in which we could analyze the transport properties. Our results rely on an atomic-scale reconstruction of the GB with the formation of structural units of approximately $3\times6$ sites extension, a property which was observed for large angle GBs. It is tempting to relate these findings to the experimental observations of an increasing GB resistance below approximately 300~K. It needs to be noted that the increase of the resistance is not as strong as in the experiments. This may result from an underestimate of the on-site repulsion ($U=2t$). However, we also expect that the formation of magnetic moments at the GB induces correlation effects for larger values of $U$ which have not been implemented in the present scheme. Specifically, the formation of singlets between nearby moments with the strongest exchange coupling and a Kondo-like screening of remaining moments is speculated to modify the temperature-dependent resistance. In fact, a distribution of Kondo temperatures $T_K$ can produce a linear resistance up to the highest value of $T_K$.~\cite{Miranda} With a measured linear resistance up to 300~K, this scenario is rather hard to implement. In a different approach, Hirsch~\cite{Hirsch} applied a scheme which builds on a `dynamic Hubbard model', an extension of the standard Hubbard model that implements the expansion of atomic orbitals upon double occupancy. He finds that the hole density near the GB increases as temperature increases. However, it is not yet obvious if this scheme will generate a linear $R(T)$ at low temperatures.

\section*{Acknowledgements}  
This work was  supported by the DFG (TRR~80)
The authors acknowledge helpful discussions with B.~M.~Andersen, U.~Eckern, S.~Graser, R.~Held, J.~E.~Hirsch, P.~J.~Hirschfeld, F.~Loder, J.~Mannhart, M.~Schmid, K.~Steffen, and F.~A.~Wolf.

\end{document}